\documentclass[aps, reprint, amsmath, amssymb, superscriptaddress, longbibliography, pra]{revtex4-2}

\usepackage{graphicx}
\usepackage{dcolumn}
\usepackage{bm}
\usepackage{amsmath}
\usepackage{amssymb}
\usepackage{cancel}
\usepackage{bbold}
\usepackage{siunitx}
\usepackage{color}
\usepackage{braket}
\usepackage{lipsum}
\usepackage{hyperref}
\usepackage{tabularx}
\usepackage{xr}
\usepackage{zref-xr}
\usepackage{lineno}
\zxrsetup{tozreflabel=false, toltxlabel=true, verbose}
\usepackage{braket}
\newcommand{\paulix}{\hat{\sigma}_\mathrm{x}}
\newcommand{\pauliy}{\hat{\sigma}_\mathrm{y}}
\newcommand{\pauliz}{\hat{\sigma}_\mathrm{z}}

\newcommand{\matindex}[1]{\mbox{\scriptsize#1}}
\usepackage{blkarray}
\zexternaldocument*{main}

\begin{document}

\setcitestyle{square}
\newcommand{\note}[1]{\textbf{#1}}

\title{A Superconducting Qubit-Resonator Quantum Processor with Effective All-to-All Connectivity}
\author{\firstname{Michael}~\surname{Renger}}\altaffiliation[M.\:R.,\:J.\:V.\:and\:N.\:W.\:contributed\:equally]{}
\email{michael.renger@meetiqm.com, jeroen.verjauw@meetiqm.com, nicola.wurz@meetiqm.com}
\affiliation{IQM Quantum Computers, Georg-Brauchle-Ring 23-25, 80992 Munich, Germany}
\author{\firstname{Jeroen}~\surname{Verjauw}}
\altaffiliation[M.\:R.,\:J.\:V.\:and\:N.\:W.\:contributed\:equally]{}
\email{michael.renger@meetiqm.com, jeroen.verjauw@meetiqm.com, nicola.wurz@meetiqm.com}
\affiliation{IQM Quantum Computers, Georg-Brauchle-Ring 23-25, 80992 Munich, Germany}
\author{\firstname{Nicola}~\surname{Wurz}}
\altaffiliation[M.\:R.,\:J.\:V.\:and\:N.\:W.\:contributed\:equally]{}
\email{michael.renger@meetiqm.com, jeroen.verjauw@meetiqm.com, nicola.wurz@meetiqm.com}
\affiliation{IQM Quantum Computers, Georg-Brauchle-Ring 23-25, 80992 Munich, Germany}
\author{\firstname{Amin}~\surname{Hosseinkhani}}
\affiliation{IQM Quantum Computers, Georg-Brauchle-Ring 23-25, 80992 Munich, Germany}
\author{\firstname{Caspar}~\surname{Ockeloen-Korppi}}
\affiliation{IQM Quantum Computers, Keilaranta 19, 02150 Espoo, Finland}
\author{\firstname{Wei}~\surname{Liu}}
\affiliation{IQM Quantum Computers, Keilaranta 19, 02150 Espoo, Finland}
\author{\firstname{Aniket}~\surname{Rath}}
\affiliation{IQM Quantum Computers, Georg-Brauchle-Ring 23-25, 80992 Munich, Germany}
\author{\firstname{Manish}~\surname{J. Thapa}}
\affiliation{IQM Quantum Computers, Georg-Brauchle-Ring 23-25, 80992 Munich, Germany}
\author{\firstname{Florian}~\surname{Vigneau}}
\affiliation{IQM Quantum Computers, Georg-Brauchle-Ring 23-25, 80992 Munich, Germany}
\author{\firstname{Elisabeth}~\surname{Wybo}}
\affiliation{IQM Quantum Computers, Georg-Brauchle-Ring 23-25, 80992 Munich, Germany}
\author{\firstname{Ville}~\surname{Bergholm}}
\affiliation{IQM Quantum Computers, Keilaranta 19, 02150 Espoo, Finland}
\author{\firstname{Chun Fai}~\surname{Chan}}
\affiliation{IQM Quantum Computers, Keilaranta 19, 02150 Espoo, Finland}
\author{\firstname{B\'alint}~\surname{Csat\'ari}}
\affiliation{IQM Quantum Computers, Keilaranta 19, 02150 Espoo, Finland}
\author{\firstname{Saga}~\surname{Dahl}}
\affiliation{IQM Quantum Computers, Keilaranta 19, 02150 Espoo, Finland}
\author{\firstname{Rakhim}~\surname{Davletkaliyev}}
\affiliation{IQM Quantum Computers, Keilaranta 19, 02150 Espoo, Finland}
\author{\firstname{Rakshyakar}~\surname{Giri}}
\affiliation{IQM Quantum Computers, Georg-Brauchle-Ring 23-25, 80992 Munich, Germany}
\author{\firstname{Daria}~\surname{Gusenkova}}
\affiliation{IQM Quantum Computers, Georg-Brauchle-Ring 23-25, 80992 Munich, Germany}
\author{\firstname{Hermanni}~\surname{Heimonen}}
\affiliation{IQM Quantum Computers, Keilaranta 19, 02150 Espoo, Finland}
\author{\firstname{Tuukka}~\surname{Hiltunen}}
\affiliation{IQM Quantum Computers, Keilaranta 19, 02150 Espoo, Finland}
\author{\firstname{Hao}~\surname{Hsu}}
\affiliation{IQM Quantum Computers, Georg-Brauchle-Ring 23-25, 80992 Munich, Germany}
\author{\firstname{Eric}~\surname{Hyypp\"a}}
\affiliation{IQM Quantum Computers, Keilaranta 19, 02150 Espoo, Finland}
\author{\firstname{Joni}~\surname{Ikonen}}
\affiliation{IQM Quantum Computers, Keilaranta 19, 02150 Espoo, Finland}
\author{\firstname{Tyler}~\surname{Jones}}
\affiliation{IQM Quantum Computers, Georg-Brauchle-Ring 23-25, 80992 Munich, Germany}
\author{\firstname{Shabeeb}~\surname{Khalid}}
\affiliation{IQM Quantum Computers, Keilaranta 19, 02150 Espoo, Finland}
\author{\firstname{Seung-Goo}~\surname{Kim}}
\affiliation{IQM Quantum Computers, Keilaranta 19, 02150 Espoo, Finland}
\author{\firstname{Miikka}~\surname{Koistinen}}
\affiliation{IQM Quantum Computers, Keilaranta 19, 02150 Espoo, Finland}
\author{\firstname{Anton}~\surname{Komlev}}
\affiliation{IQM Quantum Computers, Keilaranta 19, 02150 Espoo, Finland}
\author{\firstname{Janne}~\surname{Kotilahti}}
\affiliation{IQM Quantum Computers, Keilaranta 19, 02150 Espoo, Finland}
\author{\firstname{Vladimir}~\surname{Kukushkin}}
\affiliation{IQM Quantum Computers, Keilaranta 19, 02150 Espoo, Finland}
\author{\firstname{Julia}~\surname{Lamprich}}
\affiliation{IQM Quantum Computers, Georg-Brauchle-Ring 23-25, 80992 Munich, Germany}
\author{\firstname{Alessandro}~\surname{Landra}}
\affiliation{IQM Quantum Computers, Keilaranta 19, 02150 Espoo, Finland}
\author{\firstname{Lan-Hsuan}~\surname{Lee}}
\affiliation{IQM Quantum Computers, Keilaranta 19, 02150 Espoo, Finland}
\author{\firstname{Tianyi}~\surname{Li}}
\affiliation{IQM Quantum Computers, Keilaranta 19, 02150 Espoo, Finland}
\author{\firstname{Per}~\surname{Liebermann}}
\affiliation{IQM Quantum Computers, Keilaranta 19, 02150 Espoo, Finland}
\author{\firstname{Sourav}~\surname{Majumder}}
\affiliation{IQM Quantum Computers, Georg-Brauchle-Ring 23-25, 80992 Munich, Germany}
\author{\firstname{Janne}~\surname{M\"antyl\"a}}
\affiliation{IQM Quantum Computers, Keilaranta 19, 02150 Espoo, Finland}
\author{\firstname{Fabian}~\surname{Marxer}}
\affiliation{IQM Quantum Computers, Keilaranta 19, 02150 Espoo, Finland}
\author{\firstname{Arianne}~\surname{Meijer - van de Griend}}
\affiliation{IQM Quantum Computers, Keilaranta 19, 02150 Espoo, Finland}
\author{\firstname{Vladimir}~\surname{Milchakov}}
\affiliation{IQM Quantum Computers, Keilaranta 19, 02150 Espoo, Finland}
\author{\firstname{Jakub}~\surname{Mro\.zek}}
\affiliation{IQM Quantum Computers, Keilaranta 19, 02150 Espoo, Finland}
\author{\firstname{Jayshankar}~\surname{Nath}}
\affiliation{IQM Quantum Computers, Georg-Brauchle-Ring 23-25, 80992 Munich, Germany}
\author{\firstname{Tuure}~\surname{Orell}}
\affiliation{IQM Quantum Computers, Keilaranta 19, 02150 Espoo, Finland}
\author{\firstname{Miha}~\surname{Papi\v{c}}}
\affiliation{IQM Quantum Computers, Georg-Brauchle-Ring 23-25, 80992 Munich, Germany}
\affiliation{Department of Physics and Arnold Sommerfeld Center for Theoretical Physics, Ludwig-Maximilians-Universität München, Theresienstr.\:37, 80333 Munich, Germany}
\author{\firstname{Matti}~\surname{Partanen}}
\affiliation{IQM Quantum Computers, Keilaranta 19, 02150 Espoo, Finland}
\author{\firstname{Alexander}~\surname{Plyushch}}
\affiliation{IQM Quantum Computers, Keilaranta 19, 02150 Espoo, Finland}
\author{\firstname{Stefan}~\surname{Pogorzalek}}
\affiliation{IQM Quantum Computers, Georg-Brauchle-Ring 23-25, 80992 Munich, Germany}
\author{\firstname{Jussi}~\surname{Ritvas}}
\affiliation{IQM Quantum Computers, Keilaranta 19, 02150 Espoo, Finland}
\author{\firstname{Pedro}~\surname{Figueroa Romero}}
\affiliation{IQM Quantum Computers, Georg-Brauchle-Ring 23-25, 80992 Munich, Germany}
\author{\firstname{Ville}~\surname{Sampo}}
\affiliation{IQM Quantum Computers, Keilaranta 19, 02150 Espoo, Finland}
\author{\firstname{Marko}~\surname{Sepp\"al\"a}}
\affiliation{IQM Quantum Computers, Keilaranta 19, 02150 Espoo, Finland}
\author{\firstname{Ville}~\surname{Selinmaa}}
\affiliation{IQM Quantum Computers, Keilaranta 19, 02150 Espoo, Finland}
\author{\firstname{Linus}~\surname{Sundstr\"om}}
\affiliation{IQM Quantum Computers, Keilaranta 19, 02150 Espoo, Finland}
\author{\firstname{Ivan}~\surname{Takmakov}}
\affiliation{IQM Quantum Computers, Keilaranta 19, 02150 Espoo, Finland}
\author{\firstname{Brian}~\surname{Tarasinski}}
\affiliation{IQM Quantum Computers, Keilaranta 19, 02150 Espoo, Finland}
\author{\firstname{Jani}~\surname{Tuorila}}
\affiliation{IQM Quantum Computers, Keilaranta 19, 02150 Espoo, Finland}
\author{\firstname{Olli}~\surname{Tyrkkö}}
\affiliation{IQM Quantum Computers, Keilaranta 19, 02150 Espoo, Finland}
\author{\firstname{Alpo}~\surname{V\"alimaa}}
\affiliation{IQM Quantum Computers, Keilaranta 19, 02150 Espoo, Finland}
\author{\firstname{Jaap}~\surname{Wesdorp}}
\affiliation{IQM Quantum Computers, Keilaranta 19, 02150 Espoo, Finland}
\author{\firstname{Ping}~\surname{Yang}}
\affiliation{IQM Quantum Computers, Georg-Brauchle-Ring 23-25, 80992 Munich, Germany}
\author{\firstname{Liuqi}~\surname{Yu}}
\affiliation{IQM Quantum Computers, Keilaranta 19, 02150 Espoo, Finland}
\author{\firstname{Johannes}~\surname{Heinsoo}}
\affiliation{IQM Quantum Computers, Keilaranta 19, 02150 Espoo, Finland}
\author{\firstname{Antti}~\surname{Veps\"al\"ainen}}
\affiliation{IQM Quantum Computers, Keilaranta 19, 02150 Espoo, Finland}
\author{\firstname{William}~\surname{Kindel}}
\affiliation{IQM Quantum Computers, Georg-Brauchle-Ring 23-25, 80992 Munich, Germany}
\author{\firstname{Hsiang-Sheng}~\surname{Ku}}
\affiliation{IQM Quantum Computers, Georg-Brauchle-Ring 23-25, 80992 Munich, Germany}
\author{\firstname{Frank}~\surname{Deppe}}
\email{frank.deppe@meetiqm.com}
\affiliation{IQM Quantum Computers, Georg-Brauchle-Ring 23-25, 80992 Munich, Germany}

\date{\today}

\begin{abstract}
In this work we introduce a superconducting quantum processor architecture that uses a transmission-line resonator to implement effective all-to-all connectivity between six transmon qubits. This architecture can be used as a test-bed for algorithms that benefit from high connectivity. We show that the central resonator can be used as a computational element, which offers the flexibility to encode a qubit for quantum computation or to utilize its bosonic modes which further enables quantum simulation of bosonic systems. To operate the quantum processing unit (QPU), we develop and benchmark the qubit-resonator conditional Z gate and the qubit-resonator MOVE operation. The latter allows for transferring a quantum state between one of the peripheral qubits and the computational resonator. We benchmark the QPU performance and achieve a genuinely multi-qubit entangled Greenberger-Horne-Zeilinger (GHZ) state over all six qubits with a readout-error mitigated fidelity of $F_{\ket{\psi}_{\rm GHZ}} = 0.86$.
\end{abstract}

\maketitle
\section{Introduction}\label{sec:StarQPUintroduction}

The design flexibility of superconducting qubits facilitates the exploration of innovative quantum processing architectures with different topologies~\cite{blais2021, doi:10.1146/annurev-conmatphys-031119-050605, wendin2017, Bravyi2022}. Currently, quantum processors built from regular lattices of sparsely connected superconducting qubits have been at the center of the effort in scaling superconducting quantum processing units (QPUs) by major industrial and governmental players, with example topologies including square lattice~\cite{Acharya2024, Arute2019, Krinner2022, wu2021strong, abdurakhimov2024}, heavy hex~\cite{Hertzberg2021} and square-octagon ~\cite{Dupont2023}. In these topologies, each of the qubits is connected to at most four neighboring qubits. These QPUs are conceptually straightforward to scale, suited for applications such as the square lattice surface code~\cite{Dennis2002, Fowler2012}, and have achieved breakthroughs in science and engineering including the simulation of a Bose–Hubbard lattice~\cite{Karamlou2024} and trotterized Ising model~\cite{Kim2023}, as well as lattice surgery~\cite{lacroix2024}. However, achieving quantum utility remains elusive~\cite{Pan2022,Tindall2024,Gao2024,begusic2024}.

A large benefit of square-lattice QPUs is the possibility of algorithm execution with a high level of parallelization. On the other hand, entangling distant qubits on such a hardware platform requires the concatenation of multiple gates in SWAP networks~\cite{o2019generalized}. This leads to a gate count overhead in case the connectivity of the QPU in terms of native two-qubit gates is not matched to the algorithm of interest~\cite{Algaba2022}. Consequently, it can be beneficial to increase the connectivity between qubits at the cost of reduced parallelism for a given application. Examples for such applications include quantum error correcting codes with high encoding rates ~\cite{Bravyi2024} and noisy intermediate scale quantum (NISQ) algorithms~\cite{Kjaergaard2020}. Particularly high connectivity is provided by a star topology where all qubits are connected to a central element \cite{Xuntao2024, hazra2021ring, Malekakhlagh2024}. A promising choice for this central node is a resonator which 
has a long history in circuit QED \cite{Wallraff2004, Blais2004, Huber2024}. In particular, the combination of superconducting qubits and resonators has been employed to realize a quantum von Neumann architecture \cite{Mariantoni2011} or the resonator–zero-qubit (RezQu) scheme \cite{Galiautdinov2012}. In addition, QPUs consisting of qubits coupled to a bus resonator have demonstrated genuine multipartite entangled states for up to 20 qubits \cite{Song2017, Song2019}. Furthermore, they allow for combining  conventional qubit-based quantum computing with bosonic resonator modes that provide additional computational degrees of freedom ~\cite{Lloyd1999, Braunstein2005, Langford2017, liu2024, Leppakangas2025}. Such quantum computing modality combines the precise control of qubits with the large Hilbert space of bosonic modes. Realizing QPUs with these architectures requires engineering and mastering the control of qubit-resonator interactions and developing new methods for benchmarking.

In this paper we introduce a qubit-resonator QPU where a computational resonator is connected to six peripheral qubits in a star topology. A crucial limitation in previous work on qubit-resonator QPUs relying on passive capacitive coupling is hybridization crosstalk to the spectator qubits. One way to suppress the resulting idling error has been demonstrated in Ref.\,\cite{Galiautdinov2012} by introducing the RezQu architecture, relying on additional memory resonators for storing the quantum state of the spectator qubits during idling. In this work, we follow an alternative approach by employing frequency-tunable couplers. These active elements have been successfully employed to mediate the coupling between individual qubit-qubit pairs \cite{Chen2014, Arute2019, Li2020, Xu2020, Collodo2020, sung2021, Marxer2023, Zhang2024}. They feature a large on-off ratio in the coupling strength, while simultaneously suppressing the hybridization crosstalk.
We implement and benchmark two qubit-resonator operations, which combined with single-qubit gates, form a universal set of quantum operations for this QPU. The qubit-resonator operations are based on the Jaynes-Cummings interaction between a qubit and the central computational resonator (CR)~\cite{Blais2004}. For the MOVE operation~\cite{Galiautdinov2012} we restrict the input state of the qubit-resonator pair to the single-photon manifold and tune the qubit and the resonator into resonance in order to fully transfer a state between them. In addition, we implement a conditional Z (CZ) gate to create entanglement between a qubit and the resonator. The CZ gate is implemented by tuning the e-f transition of the qubit into resonance with the computational resonator~\cite{sung2021}. After a full Rabi cycle, the state where both the qubit and the computational resonator are in the first excited state has acquired a $\pi$ phase shift. By combining these qubit-resonator operations, we can effectively implement a CZ gate between any two qubits. We consider our system as a QPU with effective all-to-all connectivity between the qubits, because a constant overhead of two MOVE operations is required to realize a qubit-qubit CZ gate - independent of the physical distance between the qubits. Depending on the algorithm, however, it may be beneficial to treat the resonator as a proper computational element by applying MOVE and CZ operations individually. As an example for such an algorithm and to provide a global benchmark for the quantum computing capability of our QPU, we entangle all qubits in a Greenberger-Horne-Zeilinger (GHZ) state. As an outlook towards hybrid quantum computing including higher order Fock states, we demonstrate that the higher resonator levels can be populated by repeatedly exciting the qubit and applying the MOVE operation. We eventually provide a NISQ application use case for our QPU by applying error mitigation to determine the ground state energy of the transverse-field Ising model.
\section{Qubit-Resonator QPU}\label{sec:design}

\begin{figure}
    \centering
    \includegraphics[width=\linewidth]{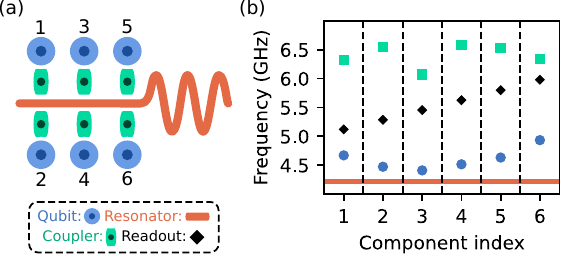}
    \caption{(a) Schematic illustration of the qubit-resonator QPU. The readout resonators (one for each qubit) have been omitted for the sake of clarity. (b) Operational frequencies of the
    qubits (blue circles), the tunable couplers (green squares),  the computational resonator (red line), and the dispersively coupled resonators used for multiplexed readout (black diamonds). The qubits and couplers are biased at their first-order flux-insensitive point and their frequencies are tunable from this maximum attainable value down to nearly zero.}
    \label{fig:topology_frequencies}
\end{figure}

The QPU consists of six frequency-tunable superconducting transmon qubits (QB)~\cite{Koch2007}, connected to a central computational resonator in a star topology, see Fig.~\ref{fig:topology_frequencies}(a). The central resonator is implemented by a co-planar waveguide transmission line in the quarter wavelength configuration. 
To achieve tunable coupling with a large on-off ratio, we employ an additional frequency-tunable transmon qubit as coupling element~\cite{sung2021,Marxer2023} between the resonator and each qubit. Analog to their use in qubit-qubit gates, these tunable couplers (TCs) mediate an effective qubit-resonator interaction, by creating an indirect coupling with a tunable sign and magnitude in addition to a small fixed qubit-resonator coupling. During gate operation, the couplers are dynamically tuned near the qubit and computational resonator frequencies, enabling fast non-adiabatic gate operation. On the other hand, in the idling configuration the total $ZZ$-coupling can be suppressed to a very small value on the order of a few kilohertz by choosing the tunable coupler frequency such that the coupler-mediated and direct couplings cancel each other. The frequencies of all the QPU components are shown in Fig.~\ref{fig:topology_frequencies}(b).

To maximize the coupling strengths, we connect the tunable couplers to the open-circuit end of the resonator. To enable flexible positioning of the qubits with respect to the computational resonator, we employ coupler extender waveguides between the components and the coupler. This modification to the tunable couplers ensures large qubit–qubit separations, thereby suppressing parasitic crosstalk~\cite{Marxer2023}.
We note that the star topology with a computational resonator as the central element can be extended beyond the six peripheral qubits shown here without fundamental limitations. We estimate that, using similar coupling values as reported here and maintaining the mm-scale spacing between neighboring qubits, up to twelve qubits can be coupled to such a single computational resonator. The qubit-resonator QPU design and fabrication methodology is discussed in Appendix~\ref{sec:app_design}.

The QPU further contains drive lines for $XY$-control of the qubits, flux bias lines to control the qubit and tunable coupler frequencies, and readout resonators that enable multiplexed dispersive readout of the qubit state. The design of the control lines and readout structures closely follows the design in Ref.~\cite{Marxer2023}.

 \section{Qubit-Resonator Operations and Characterization}\label{sec:design}

We achieve effective all-to-all connectivity on the qubit-resonator QPU with star topology by implementing CZ gates between each pair of qubits.
To realize a CZ gate CZ(QB\,2, QB\,1) between a pair of qubits, QB\,1 and QB\,2, we transfer the state from one of the qubits, i.e., QB\,1 to the computational resonator by applying MOVE(QB\,1, CR). Consecutively, we perform a CZ(QB\,2, CR) gate between the CZ qubit, QB\,2, and the resonator to entangle the two components. Then, as the last step, we apply a second MOVE operation MOVE(QB\,1, CR), which transfers the entanglement with QB\,2 from the computational resonator to the MOVE qubit, QB\,1. Hence, to create entanglement between the peripheral qubits of the device, both native qubit-resonator operations are required, the MOVE operation and the CZ gate.

The MOVE operation transfers a state from a qubit to the computational resonator or vice versa, assuming that only one of the two components is populated beforehand. We implement this operation using the resonant Jaynes-Cummings interaction described by the Hamiltonian:

\begin{align}\label{Eq:JC_int_Hamiltonian}
     \hat{H}_\mathrm{JC} = &\frac{\hbar \omega_\mathrm{QB}}{2}\hat{\sigma}_z + \hbar \omega_\mathrm{CR}\left(\hat{a}^\dagger\hat{a}+\frac{1}{2}\right) \nonumber \\
     &+ \frac{\hbar \Omega}{2}\left(\hat{a}\hat{\sigma}^+ + \hat{a}^\dagger\hat{\sigma}^-\right),
\end{align}

where $\hat{a}$ ($\hat{a}^\dagger$) is the annihilation (creation) operator for the computational resonator mode, $\hat{\sigma}^+ \equiv |e\rangle \langle g| = (\hat{\sigma}^-)^\dagger$ are the raising and lowering operators for the qubit, and $\Omega$ is the Rabi rate. For the MOVE operation, the qubit frequency, $\omega_\mathrm{QB}$, is adjusted to match the computational resonator frequency, $\omega_\mathrm{CR}$. We experimentally realize this operation by applying a magnetic flux pulse on the qubit SQUID loop.

In general, the Jaynes-Cummings interaction couples a qubit with the entire ladder of resonator number states. By integrating the time evolution of the Jaynes-Cummings interaction given in Eq.~\eqref{Eq:JC_int_Hamiltonian} for a fixed duration, we derive the unitary of the Jaynes-Cummings gate given by the transformation of the basis states
\begin{align}\label{Eq:JC_gate}
 & |g, n \rangle \rightarrow
    c_n(\Theta) |g, n \rangle - i s_n(\Theta) |e, n-1\rangle, \\ \notag
 & |e, n\rangle \rightarrow c_{n+1}(\Theta)|e, n\rangle - i s_{n+1}(\Theta) |g, n+1\rangle,
 \end{align}
where we define
\begin{equation}\label{Eq:cn_sn_definition}
c_n(\Theta) \equiv \cos\left(\frac{\sqrt{n}}{2}\Theta\right), \qquad s_n(\Theta) \equiv \sin\left(\frac{\sqrt{n}}{2}\Theta\right).
\end{equation}
Here, the angle $\frac{\Theta}{2} = \frac{1}{2}\int  \Omega(t)dt$ is the exchange angle that parametrizes the Jaynes-Cummings gate. The MOVE operation, defined for the input states $\ket{g, 1}$, $\ket{e, 0}$ and $\ket{g, 0}$ is calibrated to achieve a full population exchange of $|g, 1\rangle $ and $|e, 0 \rangle$, corresponding to an exchange angle of ${\Theta} ={\pi}$. In this case, we use the abbreviations $c_n \equiv c_n(\pi),\ s_n \equiv s_n(\pi)$.

For conventional quantum computing applications, we restrict the resonator such that it is never occupied beyond the first excited state, and thus it can be effectively mapped onto the state space of a qubit. Specifically, we carefully design the gate sequences so that the MOVE operation is never applied to a qubit-resonator state with an $|e, 1 \rangle$ contribution. As can be seen from the description of the Jaynes-Cummings gate in Eq.~\eqref{Eq:JC_gate}, the system evolves into the state $c_2 |e, 1 \rangle - i  s_2|g, 2\rangle$ if the MOVE operation is applied to $|e, 1\rangle$, thereby breaking the qubit mapping of the resonator by occupying its $\ket{2}$ state.

By avoiding the $|e, 1\rangle$ state when applying the MOVE operation, we can operate within the single photon manifold where we can conceptually treat the resonator as a qubit. By subsequently removing the restrictions on the resonator, we can access higher resonator modes. 

\subsection{QPU Characterization}
\begin{figure}
    \centering
    \includegraphics[width=\linewidth]{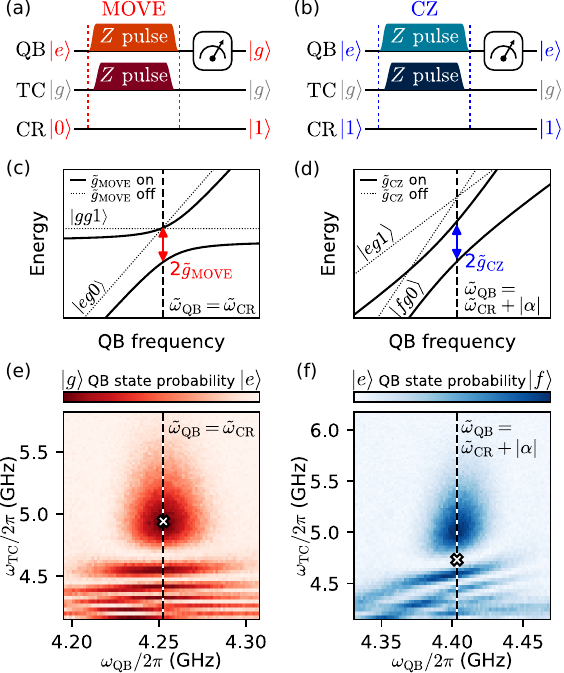}
    \caption{Qubit-resonator gate calibration. (a), (b) Pulse level schedules used for tuning up the population exchange of the MOVE operation and the CZ gate, respectively. For the MOVE operation, initially only the qubit is in the excited state, while for the CZ gate both the qubit and the computational resonator start in the first excited state. (c), (d) Energy level diagrams showing the relevant states involved in the MOVE operation and the CZ gate, respectively. Both diagrams depict the diabatic (dotted grey line) and dressed (black solid line) energy levels. The energy splittings ($2\Tilde{g}_\mathrm{MOVE},\ 2\Tilde{g}_\mathrm{CZ}$) are tuned by adjusting the coupler frequency $\omega_\mathrm{TC}$. (e), (f) Population oscillations between the energy levels shown in (c) and (d), respectively, are obtained by scanning the qubit frequency during the gate, $\omega_\mathrm{QB}$, over the resonance condition (indicated by the black dashed line), and by tuning the effective interaction strength between the qubit and the computational resonator during the gate by adjusting the coupler frequency, $\omega_\mathrm{TC}$. The white crosses indicate the initial guesses for the operating point of the gates based on the population oscillations.}
    \label{fig:2qb_gat_cal}
\end{figure}

To perform quantum operations on the qubit-resonator QPU, we employ six microwave drive lines for $XY$-control of the qubits and twelve flux lines to enable frequency tuning of the qubits and the tunable couplers with $Z$ pulses. Details of the experimental setup can be found in Appendix~\ref{sec:exp_setup}. We determine the qubit frequencies from a variable delay Ramsey experiment and extract the qubit and coupler flux dispersion by sweeping the applied DC bias. We implement single-qubit gates using the derivative removal by adiabatic gate (DRAG) method with a cosine shaped envelope of the in-phase component \cite{motzoi2009simple}. We employ error amplification techniques to reach an average individual single-qubit gate fidelity of $99.94 \pm \SI{0.09}{\percent}$ \cite{Hyyppa2024}. We characterize the coherence times of all qubits in their first-order flux-insensitive configuration. We calculate the average energy relaxation time $T_1 = 41.3 \pm \SI{1.3}{\micro \second}$, i.e., the arithmetic mean over all qubits. Using a Ramsey sequence, we find an average dephasing time of $T_2^* = 33.0 \pm \SI{1.2}{\micro \second}$. To remove the effect of quasi-static noise, we perform a Hahn echo experiment and obtain an average echo dephasing time of $T_2^\mathrm{e} = 45.8 \pm \SI{1.2}{\micro \second}$. We find an exponential decay envelope, implying white noise as dominating source of decoherence \cite{Martinis2003, Bylander2011}.

Since the computational resonator state cannot be directly prepared and read out, we measure its coherence properties indirectly by transferring a state between one of the qubits and the resonator using the MOVE operation. For the resonator relaxation time, we obtain  $T_1 = 5.53\pm\SI{0.32}{\micro \second}$. 
Via a Ramsey sequence, we measure a resonator dephasing time of $T_2^* = 10.9\pm\SI{1.0}{\micro \second}$ close to $2T_1$, implying that the resonator coherence is limited by energy relaxation. 
We furthermore extract the computational resonator frequency at the idling configuration, $\omega_\mathrm{CR}=\SI{4.218}{\giga \hertz}$. 
The resonator relaxation is one of the limiting factors of the presented QPU performance. Ongoing efforts to reduce the Purcell decay and the energy participation ratio \cite{Minev2021}, in addition to employing flip-chip architecture, are expected to yield significant improvements in the $T_1$ time of the computational resonator.
Details about the measurement techniques for the resonator characterization as well as a more detailed discussion about the main factors impacting the relaxation time and possible future advancements are presented in Appendix~\ref{sec:app_cr_char}. A summary of the QPU parameters can be found in Appendix~\ref{sec:qpu_charact_appendix}.

\begin{figure*}
    \centering
    \includegraphics[width=\textwidth]{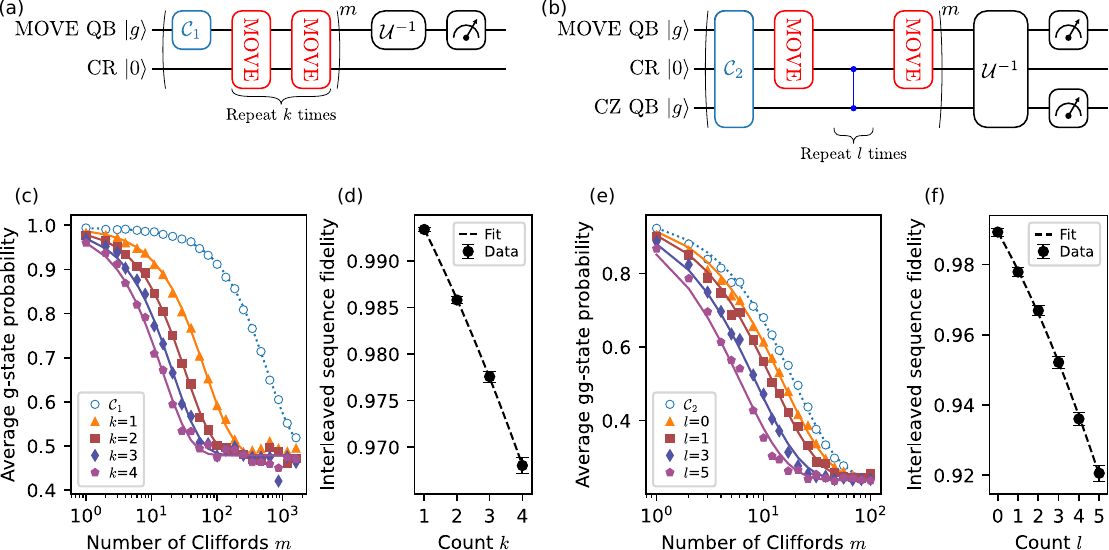}
    \caption{Interleaved randomized benchmarking experiments for
qubit-resonator gates. (a) Circuit for benchmarking the MOVE operation via interleaved randomized benchmarking. The sequence consists of $m$ single-qubit Clifford gates $\mathcal{C}_1$ interleaved with $k$ repetitions of two subsequent MOVE operations, which effectively implement an identity. The final operation $\mathcal{U}^{-1}$ inverts the applied sequence. (b) Circuit for benchmarking the CZ gate via interleaved randomized benchmarking. Two-qubit Clifford gates $\mathcal{C}_2$ are interleaved with MOVE-$l$CZ-MOVE operations for a varying number of CZ gates $l$. (c) Experimental data of the sequence fidelity for
randomized benchmarking with $k=1, .., 4$ interleaved double MOVE operations and the reference single-qubit Cliﬀord sequence (open blue circles) as
a function of the number of Cliﬀord gates. (d) Extracted fidelity of the interleaved gate sequence as a function of $k$ (solid black circles) and quadratic fit (dashed line). (e) Experimental data of the sequence fidelity for
randomized benchmarking with interleaved MOVE-$l$CZ-MOVE gate sequences and the reference two-qubit Cliﬀord sequence (open blue circles) as
a function of the number of Cliﬀord gates. (f) Extracted fidelity of the interleaved MOVE-$l$CZ-MOVE gate sequence as a function of $l$ (solid black circles) and quadratic fit (dashed line).}
    \label{fig:2qb_gate_bench}
\end{figure*}

\subsection{Calibration of Qubit-Resonator Gates}
In the following, we discuss the calibration of the qubit-resonator MOVE operation and the qubit-resonator CZ gate.

We use the notation $\ket{\mathrm{QB},\mathrm{TC},\mathrm{CR}}$ to represent the eigenstates of the system (with $\ket{g}$, $\ket{e}$, $\ket{f}$ for qubit and coupler eigenstates, and $\ket{0}$, $\ket{1}$ as resonator number states), which approximate the diabatic eigenstates of the system, i.e., at the idling configuration. The idling configuration is a chosen set of qubit, coupler and resonator frequencies which results in a minimal residual $ZZ$-interaction. 
We implement the qubit-resonator gates using
non-adiabatic transitions between $\ket{eg0}$ and $\ket{gg1}$ for the MOVE operation, and $\ket{eg1}$ and $\ket{fg0}$ for the CZ gate, see Fig.~\ref{fig:2qb_gat_cal}(a) and Fig.~\ref{fig:2qb_gat_cal}(b). This population exchange is initiated by first bringing the involved qubit-resonator states into resonance and then quickly tuning the frequency of the coupler to increase the interaction strength $\Tilde{g}_\mathrm{MOVE}$ between $\ket{eg0}$ and $\ket{gg1}$, or $\Tilde{g}_\mathrm{CZ}$  between $\ket{eg1}$ and $\ket{fg0}$, respectively.
As shown in Fig.~\ref{fig:2qb_gat_cal}(c) and Fig.~\ref{fig:2qb_gat_cal}(d), the interaction between the involved qubit-resonator states opens up an avoided crossing with an energy gap of $2\Tilde{g}_\mathrm{MOVE}$ or $2\Tilde{g}_\mathrm{CZ}$, respectively, and shifts the energy levels downwards due to the dispersive interaction with the coupler.
For the MOVE operation, the eigenenergies of the dressed states fulfill $\Tilde{\omega}_\mathrm{QB} = \Tilde{\omega}_\mathrm{CR}$. This resonance condition differs from the bare frequencies and is the result of different hybridizations between the qubit, coupler and resonator as illustrated in Appendix~\ref{sec:SWtrans}. 
The operating point of the MOVE operation is determined from a measurement of the population exchange as shown in Fig.~\ref{fig:2qb_gat_cal}(e). For the CZ gate, we tune the $\ket{eg1}$ and $\ket{fg0}$ states close to resonance such that $\Tilde{\omega}_\mathrm{QB} = \Tilde{\omega}_\mathrm{CR} + |\alpha|$. Here $\alpha$ is the anharmonicity of the qubit participating in the CZ gate. Similarly to the MOVE, we measure the  population exchange of the CZ qubit as shown in Fig.~\ref{fig:2qb_gat_cal}(f). In contrast to the above, however, the operating point of the CZ gate is where the qubit, starting in $\ket{e}$, completes a \emph{full} oscillation through $\ket{f}$ and returns back to the initial $\ket{e}$ state, as the CZ gate only modifies the phase and not the population of the basis states. The full tune-up procedure for both qubit-resonator gates, including the calibration of the conditional $\pi$ phase shift required for the CZ gate, is discussed in detail in  Appendix~\ref{sec:cal_bench_appendix}. A derivation of the analytical expression for the conditional phase is presented in Appendix~\ref{sec:SWtrans}. 

During the qubit-resonator gates, the transition frequencies of both components change, resulting in the accumulation of single-qubit phases. We correct for these phases by applying virtual Z (VZ) gates to the respective $XY$ qubit drive pulses \cite{mckay2017efficient}.
To obtain the optimal VZ correction for the qubit when applying the MOVE operation, we perform the operation an even number of times and calibrate the common phase obtained when moving a superposition state from the qubit to the computational resonator and back. For the CZ gate, two separate VZ corrections are calibrated, one for the computational resonator and one for the CZ qubit. The VZ correction of the computational resonator is determined indirectly via the qubit that was used to prepare the state in the resonator (MOVE qubit). The relative phase rotation of a state that has been transferred to the computational resonator with respect to the rotating frame of the MOVE qubit results in a time-dependent phase that accumulates at a rate corresponding to the frequency difference between the MOVE qubit and the resonator \cite{bao2022fluxonium}. To properly correct for single-qubit phase accumulation in arbitrary algorithms, we need to distinguish this time-dependent phase from fixed phase corrections that depend only on the flux pulse shape implementing a certain gate.
Details on the calibration of the single-qubit phase corrections can be found in Appendix~\ref{Sec:MOVE_fine_VZ} and Appendix~\ref{Sec:CZ_fine_VZ}. 

\subsection{Benchmarking Qubit-Resonator Gates}\label{sec:qr_gate_characterization}

For benchmarking qubit-resonator gates, we employ the technique of interleaved randomized benchmarking (iRB) \cite{Magesan2012, Knill2008}. The average gate fidelity is estimated from the decay of the measured qubit population as a function of the applied sequence length $m$.
The iRB circuit for the MOVE operation is shown in Fig.~\ref{fig:2qb_gate_bench}(a). Since we apply the MOVE operation only to states for which either the qubit or the computational resonator are unpopulated, the problem of calculating the average gate fidelity reduces from an integration over all possible input states for a bipartite quantum system to considering only a single effective two-level system. Consequently, we benchmark the MOVE operation in the framework of single-qubit RB interleaving pairs of MOVE operations, which effectively implement an identity gate. 
To reduce the uncertainty of the extracted double MOVE fidelity, we perform multiple instances of the iRB experiment, interleaving the effective identity gate $(\mathrm{MOVE})^{2k}$ for different integer numbers $k$. The ground state probability of the qubit is then fitted by $P_k(m) = A p_k^m + B$, where the depolarization probability $p_k$ is treated as independent fit parameter for each $k$, and the amplitude $A$ as well as the offset $B$ are fitted simultaneously for all $k$. For each dedicated $m$ in the iRB experiment, we apply 60 different random Clifford sequences and average the experiment over $256$ repetitions. In Fig.~\ref{fig:2qb_gate_bench}(c), we show the result of such an iRB experiment for QB\,2 for up to $k = 4$. We extract the fidelity per interleaved gate sequence as a function of $k$, as shown in Fig.~\ref{fig:2qb_gate_bench}(d). We observe that the interleaved fidelity can be approximated by the fitting function $f_\mathrm{m}(k) = 1 - \alpha_\mathrm{m} k^2 - \beta_\mathrm{m} k$ \cite{Xiong2025}, where the quadratic contribution results from quasi-static noise and remaining gate calibration errors, and the linear term corresponds to depolarization errors. For example, a small error in the exchange angle of the MOVE operation would subsequently lead to populating higher resonator modes with increasing $k$, according to Eq.~\eqref{Eq:JC_gate}, and the loss in qubit population is, to first order, proportional to $k^2$. From the fit, we obtain $\alpha_\mathrm{m} = 4.6\times10^{-4}$ and $\beta_\mathrm{m} = 6.1\times10^{-3}$. We determine the fidelity of the double MOVE operation as $F_\mathrm{mm} = f_\mathrm{m}(1) = 99.34 \pm 0.03$. The single-qubit Clifford fidelity, averaged over all six qubits, is $0.9990\pm0.0002$ and the double MOVE fidelity averaged over all qubit-resonator pairs, is $\bar{F}_\mathrm{mm} = 0.988 \pm 0.006$, where the error bars represent the standard deviation of the respective individual fidelities.
We note that recent work has introduced network benchmarking~\cite{helsen2023benchmarking} and demonstrated its utility in evaluating state transfer operations in the context of quantum networks~\cite{heya2025randomized}.
However, this technique requires direct control over the state of all computational components, which is not feasible in our device due to the inability to apply single-qubit gates to the computational resonator.

Figure~\ref{fig:2qb_gate_bench}(b) shows the iRB circuit for estimating the qubit-resonator CZ gate fidelity. In order to sample all possible input states for the CZ gate acting between a qubit and a computational resonator, we use the MOVE operation to populate the resonator with an arbitrary state. After applying the CZ gate, the resonator has to be brought back to its vacuum state by applying another MOVE operation. In contrast to the benchmarking of the MOVE operation, here we can use two-qubit Clifford benchmarking and integrate over all possible two-qubit states of the MOVE and CZ qubit. We generate two-qubit Clifford gates from single-qubit gates and effective CZ gates acting between the MOVE and the CZ qubit, which are composed of the native qubit-resonator MOVE and CZ operations. Similar as in iterative iRB \cite{Sheldon2016}, we vary the number $l$ of interleaved CZ gates between the two MOVE operations and refer to this experiment as MOVE-$l$CZ-MOVE iRB. In Fig.~\ref{fig:2qb_gate_bench}(e) we show the corresponding iRB result when using QB\,1 as the CZ qubit and QB\,2 as the MOVE qubit, for up to $l=5$ CZ gates. The data is obtained using 30 different random Clifford sequences and averaged over $256$ repetitions. We extract the interleaved fidelity as a function of $l$ as shown in Fig.~\ref{fig:2qb_gate_bench}(f). Notably, this quantity reflects not only the performance of the qubit-resonator CZ gate but also the fidelity of the state transfer between the MOVE qubit and the computational resonator, CR, since this is a crucial prerequisite for entangling non-adjacent qubits using this gate sequence for odd values of $l$. The interleaved fidelity follows $f_\mathrm{cz}(l) = \gamma_\mathrm{m} -\alpha_\mathrm{cz} l^2 - \beta_\mathrm{cz} l$, in analogy to $f_\mathrm{m}(k)$. Here, the offset $\gamma_\mathrm{m}$ quantifies the errors induced by the MOVE operation at the beginning and at the end of the interleaved sequence. We note that the data point at $l=0$ follows the quadratic trend for $l\geq1$, therefore we can assign individual errors to the MOVE and CZ operations, even though they do not commute. More generally, the interleaved sequence fidelity exhibits a consistent trend for both even and odd values of $l$, which correspond to benchmarking an identity and CZ gate unitary, respectively. From the quadratic fit shown in Fig.~\ref{fig:2qb_gate_bench}(f), we obtain $\alpha_\mathrm{cz} = 7.32\times10^{-4}$, $\beta_\mathrm{cz} = 1.01\times 10^{-2}$ and $\gamma_\mathrm{m} = 0.989$. We observe that $\gamma_\mathrm{m}$ is slightly smaller than the double MOVE fidelity, $F_\mathrm{mm}$, extracted from the single-qubit RB presented above. We attribute the mismatch in the fidelity to the idling infidelity of the CZ qubit during the two MOVE operations, which only affects $\gamma_\mathrm{m}$. Furthermore, the input state sampling in two-qubit Clifford benchmarking is different as compared to single-qubit RB, which can slightly affect the inferred double MOVE fidelity. The fidelity of the qubit-qubit CZ gate, implemented as a sequence of MOVE-CZ-MOVE gates, is directly measured using the iRB protocol for $l=1$ and the result is $0.978 \pm 0.002$. The corresponding two-qubit Clifford fidelity is $0.9625 \pm 0.0007$. We extract the fidelity of an individual CZ gate between QB\,1 and the computational resonator as $F_\mathrm{cz}=f_\mathrm{cz}(1)/\gamma_\mathrm{m}=98.90 \pm 0.05$. 
Because the value of the quadratic coefficient, $\alpha_\mathrm{cz}$, is small compared to the linear coefficient, $\beta_\mathrm{cz}$, we conclude that the depolarization error is the main mechanism limiting the CZ gate fidelity.  Furthermore, we average the fidelity of all six qubit-resonator CZ gates and obtain $\bar{F}_\mathrm{cz} = 0.983 \pm 0.009$, where the error bar represents the standard deviation over the respective individual gate fidelities. The individual double MOVE and CZ fidelities are listed in Appendix~\ref{sec:qpu_charact_appendix}.

Next, we estimate the fidelity limits set by decoherence \cite{Abad2022}. For the CZ gate, we additionally take into account that we leave the computational space during the $|1,e\rangle \leftrightarrow |2,g\rangle$ population exchange cycle \cite{Abad2024}. Details about the coherence limit estimation are provided in Appendix~\ref{sec:coherence_limit}. An estimate for the coherence limit averaged over all qubit-resonator pairs is $\bar{F}^\mathrm{c}_\mathrm{mm} = 0.993\pm 0.001$ for the double MOVE operation, and $\bar{F}^\mathrm{c}_\mathrm{cz} = 0.993\pm 0.002$ for the CZ gate. These values provide an upper bound for the coherence-limited fidelity because the coherence times used in the calculation are measured at the qubit idling frequency configuration, while there is increased flux sensitivity due to qubit frequency tuning during the gate operation. In addition, we do not take hybridization with the coupler into account. Consequently, at least $\SI{58.3}{\percent}$ ($\SI{41.2}{\percent}$) of the double MOVE (CZ) infidelity are attributed to decoherence during the operation. 
\section{Qubit-resonator QPU Applications}\label{sec:higherlevels}

We demonstrate the capabilities of our qubit-resonator QPU and inspire further algorithmic development by presenting four distinct examples. First, we investigate the generation of multipartite entanglement by preparing GHZ states. Next, we use the Q-score benchmark to determine the ability of our QPU to
solve combinatorial optimization problems. Finally, we discuss how the MOVE operation can be used as a true Jaynes-Cummings gate to access the higher levels of the computational resonator and present a simulation of the ground state energy of the transverse field Ising model using a variational approach.

\begin{figure}
    \centering
    \includegraphics[width=\linewidth]{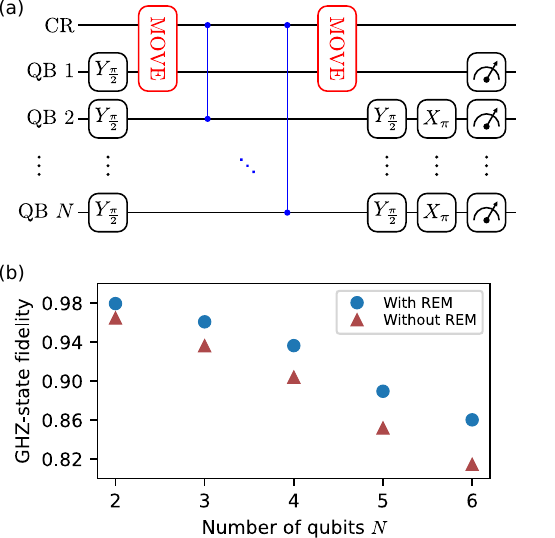}
    \caption{GHZ state characterization. (a) Circuit for preparing a GHZ state with a variable number of entangled qubits $N$ on a qubit-resonator QPU with star topology. (b) The fidelity $F_{\ket{\psi}_{\rm GHZ}}$ is calculated from the GHZ state population and its coherence, inferred via multiple quantum coherences. We show the GHZ state fidelity as a function of the qubit number $N$, obtained by analysing the data both with (blue circles) and without (red triangles) applying readout-error mitigation.}
    \label{fig:ghz_bench}
\end{figure}

\subsection{GHZ State Preparation}\label{subsec:GHZ}

As a global benchmark for the QPU performance, we examine its quantum entanglement properties. A fundamental criterion for quantum computing is the ability to entangle all qubits in the processor into a genuinely multi-qubit entangled (GME) state. Here we demonstrate GME by preparing an $n$-qubit GHZ state~\cite{greenberger1989going}, which is defined as \(\ket{\psi}_{\rm GHZ} = \frac{\ket{g}^{\otimes n} + \ket{e}^{\otimes n}}{\sqrt{2}}\). We generate the GHZ state by employing two MOVE operations between a single qubit and the computational resonator and an interleaved cascade of qubit-resonator CZ gates, see Fig.~\ref{fig:ghz_bench}(a). In between, multiple CZ gates are applied to entangle the computational resonator with all the peripheral qubits (apart from the MOVE qubit). Therefore, the only overhead in the number of gates as compared to the case of a square grid topology, are the two MOVE operations, which are required to (de)populate the computational resonator. Note that no single-qubit gates are applied to the MOVE qubit (QB\,1 in Fig.~\ref{fig:ghz_bench}(a)) between the first and second MOVE operation. This ensures that this qubit is in the ground state when the second MOVE operation is applied.

A quantum state $\ket{\psi}$ is declared to be GME if the fidelity of the experimental state $\rho$ with respect to the pure state is $F_{\ket{\psi}} = \bra{\psi} \rho \ket{\psi} >0.5$~\cite{wei2003geometric}. 
We estimate the fidelity of the prepared GHZ state using the multiple quantum coherences method~\cite{wei2020verifying, baum1985multiple} to certify its entanglement properties.

Figure~\ref{fig:ghz_bench}(b) presents GHZ state fidelities both with and without readout-error mitigation (REM). 
The results successfully confirm the presence of GME for all six qubits, demonstrating that the entire device is entangled. Without mitigation, we obtain a GHZ state fidelity of $F_{\ket{\psi}_{\rm GHZ}} = 0.815$. Furthermore, we mitigate readout errors by multiplying the raw readout results by the inverse of the readout assignment matrix, which includes both state preparation and measurement errors. The assignment matrix is a tensor product of all single-qubit assignment matrices and therefore assumes the readout errors to be completely uncorrelated. The mitigated GHZ fidelity for $N=6$ qubits is $F_{\ket{\psi}_{\rm GHZ}} = 0.86$.

\subsection{Q-score Benchmark}

In this section we assess the performance of the qubit-resonator QPU in solving combinatorial optimization problems. Specifically, we benchmark its ability to compute solutions for the maximum-cut problem using the Q-score benchmark~\cite{martiel_2011}. 
The Q-score is defined for random \textit{Erdős–Rényi graphs}, where each edge is included with probability \(1/2\). A quantum processor passes the Q-score test for a given problem size \( n \) if it achieves a performance metric \( \beta(n) \geq \beta^{\star} \), where \( \beta(n) \) quantifies the fraction of the optimality gap between a random guess and the optimal solution that is measured on the quantum system on average. If $\beta(n)= 1$, exact solutions are found, but if $\beta(n)= 0$, random solutions are found. Following Ref.~\cite{martiel_2011}, we take $\beta^{\star}=0.2$ meaning that the system passes the Q-score test for a given $n$ if it performs on average 20\% better than random guessing. Due to the accumulation of noise, passing the Q-score test becomes increasingly challenging as the problem size grows.

\begin{figure}
    \centering    
    \includegraphics[width=\linewidth]{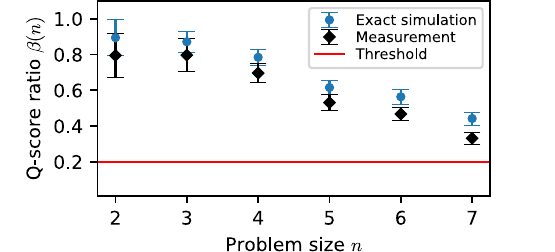}
    \caption{Measured Q-score ratio as a function of problem size $n$. For each problem size, the average is taken over 60 randomly sampled Erdős–Rényi graphs. The error bars show the standard error of the mean. }
    \label{fig:qscore}
\end{figure}

We evaluate Q-score performance $\beta(n)$ by using a depth \( p = 1 \) quantum approximate optimization algorithm (QAOA) circuit~\cite{farhi2014quantum,bib:RevModPhys.94.015004}, where the optimal circuit parameters are determined through analytical calculations~\cite{Ozaeta2020}. By ensuring that the quantum processor operates with optimally pre-tuned parameters, we avoid expensive quantum-classical optimization loops and ensure the most efficient use of quantum resources. In the QAOA algorithm, the number of RZZ interactions equals the number of edges $|E|$ in the problem graph. As we consider random Erdős–Rényi graphs with edge probability 1/2, we would need to implement on average $n(n-1)/2$ CZ gates on an all-to-all connected topology. We note that on a square-grid topology, we would require 50\% to 75\% more CZ gates for large $n$ depending on which SWAP strategy is used~\cite{Weidenfeller2022}. This highlights the potential of the star topology in solving dense optimization tasks.

To further enhance performance, we execute quantum circuits on an optimized qubit layout, following the methodology introduced in Ref.~\cite{nation2023suppressing}. This pre-processing step evaluates a noise score for each possible qubit layout, considering the gate count of the circuit along with the single- and two-qubit gate error rates of the hardware. Furthermore, we apply readout-error mitigation using the \textit{mthree package}~\cite{nation2021scalable}, allowing us to exclude measurement errors from the noise score evaluation.

We demonstrate that our QPU passes the Q-score test for problem sizes defined across all $n=6$  qubits of the QPU, as illustrated in Fig.~\ref{fig:qscore}. Additionally, we increase the Q-score to $n+1=7$ by using the virtual node technique~\cite{bravyi2020obstacles,ronkko2024}, which breaks the $\mathbb{Z}_2$-symmetry of the maximum-cut problem and introduces only a minor overhead from a few single-qubit \( Z \)-gates.

We note that for such small problem sizes, the CZ gate-count advantage of the star topology is rather small: e.g.,\ for $n=6$ we require on average 15 CZ gates on the star topology and 17 CZ gates on a square grid topology. In addition, we also need to implement $2(n-1)$ MOVE operations on the qubit-resonator star. For larger $n$, however, the gate-count advantage of the star topology becomes significant and approaches the asymptotic percentages stated above. Therefore, scaling up the qubit-resonator star topology is a worthwhile effort.

\subsection{Jaynes-Cummings Gate to Access the Higher Resonator Levels}\label{sec:higherlevels}
\begin{figure*}
	\begin{center}
		\includegraphics[width=\linewidth,angle=0,clip]{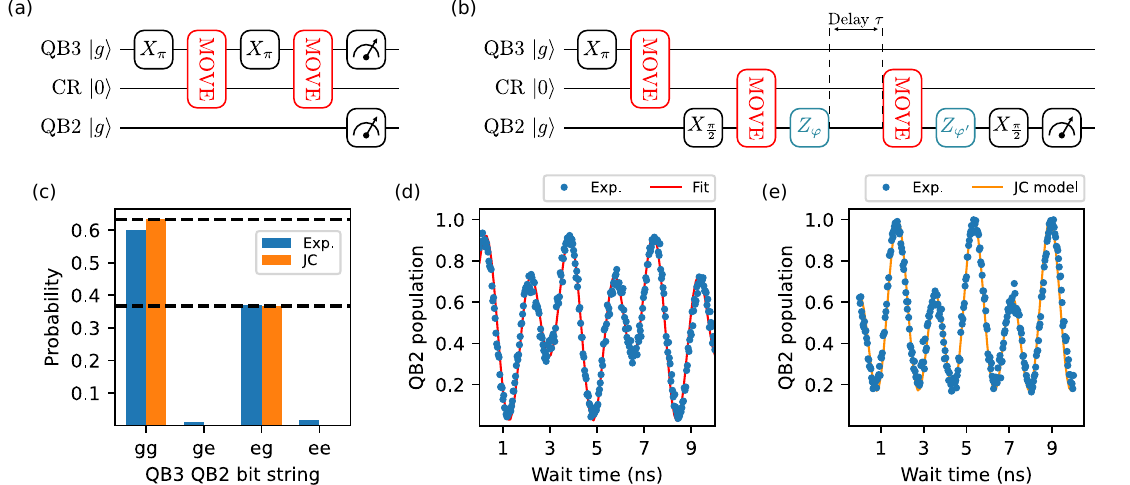}
	\end{center}
	\caption{(a) Quantum circuit to populate higher excited levels of the computational resonator by repeated excitation of QB\,3, followed by application of the MOVE operation. In addition, a second idle qubit QB\,2 is read out. (b) Circuit for measuring the free time evolution of the population of the first two excited computational resonator (CR) levels. After populating CR using QB\,3, QB\,2 is used to implement a Ramsey sequence which enables the measurement of the free resonator state evolution as a function of the delay time $\tau$. The protocol additionally contains two physical $Z$-rotations, $Z_\varphi$ and $Z_{\varphi^\prime}$, which are used to compensate for phase offsets induced by the JC interaction. (c) Detected population of QB\,3 and QB\,2 (blue) for the circuit shown in panel (a) and comparison with the ideal JC model (orange). (d) Detected population of QB\,2 as a function of $\tau$ (blue dots) for $\varphi = \varphi^\prime = 0$. The red line corresponds to a fit according to Eq.~\eqref{Eq:JC_probability_with_phase_shift}. (e) QB\,2 population as a function of $\tau$ (blue dots) after compensating all the phase offsets by employing physical $Z$-rotations $\varphi$ and  $\varphi^\prime$. The orange line is the theoretical prediction from the ideal JC model.}
	\label{Fig:Fig_JC}
\end{figure*}

In the previous section, we have restricted the MOVE operation to the single-excitation manifold of the coupled qubit-resonator system to ensure that the resonator remains in the $\{\ket{0},\ket{1}\}$ subspace. Here, we remove this restriction and populate higher resonator levels by applying the MOVE operation on the $|1,e\rangle$ state to verify that we are indeed implementing a Jaynes-Cummings gate as defined in Eq.~\eqref{Eq:JC_gate}. We construct circuits to investigate the dynamics of multiple resonator excitations $\ket{n}$ interfering with each other, similar as in Ref.\,\cite{Hofheinz2009}. With these interference fringes we measure and then correct non-ideal phases introduced by the MOVE operation to recover the dynamics of an ideal Jaynes-Cummings gate.

We can populate higher levels of the computational resonator by repeatedly employing the MOVE operation, thus providing additional computational degrees of freedom. In fact, exploiting the equidistant resonator energy spectrum can be used, for example, for digital quantum simulation \cite{Langford2017} or for the implementation of the quantum Fourier transformation in oscillating modes. In the remainder of this section, we experimentally demonstrate how higher energy levels of the computational resonator can be accessed by applying successive MOVE operations, and we provide an exemplary circuit for phase calibration.

The MOVE operation is calibrated in the single photon manifold. For states containing $n$ resonator photons, Eq.~\eqref{Eq:JC_gate} predicts that the effective Rabi frequency increases characteristically as $\Omega_n = \sqrt{n}\Omega$ \cite{Fink2008}, implying swap angles exceeding $\pi/2$ and thus creation of superposition states containing $n \geq 2$ photons. In this case, the JC interaction implements a mapping analogous to a fermionic simulation (fSim) gate \cite{Arute2019, Wang011}. If the interaction is switched on for a gate time $\tau$, the propagator corresponding to Eq.~\eqref{Eq:JC_int_Hamiltonian} leads to the general JC gate
\begin{align}\label{Eq:general_JC_interaction}
 & |g,n \rangle \rightarrow c_n^{-} |g,n \rangle - i s_n^{+} | e,n-1 \rangle, \\ \notag
 & |e,n\rangle \rightarrow c_{n+1}^{+}|e,n \rangle - i s_{n+1}^{-}|g, n+1\rangle,
\end{align}
where
\begin{align}\label{Eq:definition_cnpm_snpm}
& c_n^{\pm} \equiv e^{-i(\gamma_n \pm \zeta_n)}\cos(\Omega_n \tau), \\ \notag
& s_n^{\pm} \equiv e^{-i(\gamma_n \pm \chi_n)}\sin(\Omega_n \tau).
\end{align}
The phases $\gamma_n, \zeta_n $ and $\chi_n$ are single-qubit phases resulting from the dynamic frequency change of the involved components \cite{Arute2020}. The calibration of the MOVE operation sets the constraint $\Omega \tau = \pi/2$.

A simple circuit for demonstrating the JC interaction by subsequent application of two MOVE operations between QB\,3 and CR is shown in Fig.~\ref{Fig:Fig_JC}(a). A second idle qubit, QB\,2, is read out to test for potential cross-talk and state leakage. The qubit-qubit population as well as the theoretically expected probabilities $P_\mathrm{ee} = |s_2^{\pm}|^2 \simeq 0.63$ and $P_\mathrm{eg} = |c_2^{\pm}|^2 \simeq 0.37$ are plotted in Fig.~\ref{Fig:Fig_JC}(c). We observe that the experimental data can be well explained by climbing the JC ladder, according to Eq.~\eqref{Eq:general_JC_interaction}. 

The aforementioned $|e,1\rangle \leftrightarrow |g,2\rangle$ population exchange experiment is independent of potential phase shifts in the two-photon manifold. This is not generally the case for circuits involving superposition states or excitation swaps between the computational resonator and different qubits. In a more general case involving superposition of states containing different resonator photon number or in case of JC interaction between the resonator and multiple qubits, the phases $\gamma_n, \zeta_n$ and  $\chi_n$ need to be determined accurately, e.g., by the use of physical $Z$-rotations or Floquet theory \cite{Arute2020}. These phase shifts depend on the flux pulse window and are required to be calibrated separately for each JC manifold to realize the ideal JC interaction. For the conventional MOVE operation, the only relevant phase, $\gamma_1$, is compensated by employing a VZ rotation for the involved qubit drive signal. Correcting for these phase offsets in circuits leading to photon number contributions $n \geq 2$ requires additional control degrees of freedom such as physical $Z$-rotations. 

An example for such a circuit is shown in Fig.~\ref{Fig:Fig_JC}(b). First, we move an excitation into the computational resonator using QB\,3. Subsequently, we use QB\,2 to implement a Ramsey sequence for the populated resonator. After the first (second) MOVE operation involving QB\,2, we insert a physical $Z$-rotation $Z_\varphi$ ($Z_{\varphi^\prime}$). The frequency detuning $\Delta$ between QB\,2 and CR induces a change of the rotating reference frame when the MOVE operation is applied. This change of reference frame induces a phase evolution proportional to the delay $\tau$, manifesting as Ramsey oscillations. For the circuit in Fig.~\ref{Fig:Fig_JC}(b), the QB\,2 population can be expressed as
\begin{equation}\label{Eq:JC_probability_with_phase_shift}
P_e = \frac{1}{2}\left[1 - c_2^2 \cos(\Delta t + \phi) + s_2^2 \cos(2\Delta t + \phi^\prime) \right],
\end{equation}
where $\phi \equiv  -2\gamma_2 - 2\zeta_2 +\varphi^\prime$ and $\phi^\prime \equiv -2\gamma_2 + \varphi^\prime - \varphi$. Consequently, in contrast to the conventional Ramsey experiment with an oscillation frequency $\Delta$, we expect additional beating with frequency $2\Delta$ due to the computational resonator state contributions from $n=2$ during the free state evolution. Figure~\ref{Fig:Fig_JC}(d) shows the experimental result for the populated-resonator Ramsey experiment for detuning $\Delta/2\pi = \SI{281}{\mega \hertz}$ and $\varphi = \varphi^\prime = 0$, as well as a fit according to Eq.~\eqref{Eq:JC_probability_with_phase_shift}. From the two fit parameters, $\phi = -2.73$ and $\phi^\prime = -1.18$, we can determine the angles $\gamma_2$ and $\zeta_2$ as well as the correct choice of $\varphi$ and $\varphi^\prime$ to implement an ideal JC gate. The result from a subsequent experiment after employing the physical $Z$-corrections is shown in Fig.~\ref{Fig:Fig_JC}(e). Following this phase calibration, the circuit output can be well described by the ideal JC theory. These proof-of-principle results demonstrate that the JC gate phases can be controlled using physical $Z$-rotations. Additional information as well as the derivation of Eq.~\eqref{Eq:JC_probability_with_phase_shift} is provided in Appendix~\ref{JC_gate_appendix}.

Note that while restricting the maximal resonator photon number to one enables effective all-to-all connectivity via the MOVE operation, releasing this constraint leads to an alternative operation regime with all peripheral qubits coupled to a central active multilevel element. Consequently, our quantum hardware can be operated in two fundamentally different ways depending on a given application using the same native gate set.
\subsection{Error Mitigation to Improve Circuit Execution Reliability}\label{sec:em}

The presence of noise in near-term quantum devices introduces significant errors in measured expectation values, limiting the accuracy of quantum computations. Error mitigation techniques provide a means to recover meaningful results without requiring additional hardware resources, making them a crucial component of practical quantum computation. We demonstrate that the presented QPU is fully compatible with the state-of-the-art and established error mitigation methods, which can substantially diminish the impact of the hardware noise on the quantum algorithm execution.

We focus on estimating the ground state energy of the transverse field Ising model (TFIM), a well-known system in quantum many-body physics.  We consider this system in its critical phase (associated with a transverse field strength of $g=1$) for a six-qubit system with periodic boundary conditions and nearest-neighbor connectivity. The Hamiltonian of the model is given by
\begin{align}
\label{eq:tfi_hamiltonian}
    H = -g\sum_{j}\sigma_X^j - \sum_{(i,j)}\sigma_Z^i\sigma_Z^{j},
\end{align}
where the summation over $(i, j)$ accounts for qubit pairs with direct interactions.
\begin{figure}[t]
  \centering
  \includegraphics[width=\linewidth]{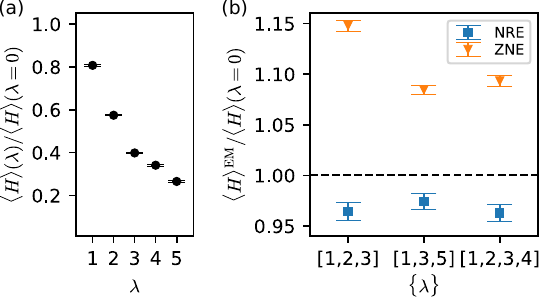}
  \caption{(a) Measured ground-state energy of the TFIM, relative to its noise-free value, as a function of the noise scale factor $\lambda$. (b) Error-mitigated ground-state energy of the TFIM, relative to its noise-free value, obtained using NRE and ZNE for different choices of noise scale factors. In the case of ZNE, an exponential fitting procedure is employed. NRE is performed following the same approach outlined in \cite{Amin2025}. Error bars for all error mitigated data points represent statistical uncertainties associated with a total of 240,000 measurement shots. }
  \label{fig:TFIM_noisy_expts}
\end{figure}

We estimate the ground-state energy with a variational approach based on QAOA with  $p=3$  layers~\cite{Wen2018}. The circuit parameters are optimized in a noise-free setting and subsequently transpiled to match the connectivity constraints and native gate set of the QPU. The resulting target circuit consists of 36 entangling CZ gates and 36 MOVE operations, which introduce noise and lead to errors in the measured energy expectation values.

To mitigate these errors, we employ a recently introduced algorithmic error mitigation technique known as noise-robust estimation (NRE) \cite{Amin2025}. This method has been demonstrated to effectively recover noise-free expectation values with high accuracy, even without applying additional error-reduction techniques such as readout error mitigation or randomized compiling \cite{Akel_2021}. For comparison, we also apply the well-established zero-noise extrapolation (ZNE) method \cite{IBM_ExpZNE2019}.

Both NRE and ZNE require amplification of hardware noise, necessitating the execution of the target circuit at different noise levels. Assuming that the circuit noise rate can be characterized by  $\epsilon_0$, we parameterize the noise as $\epsilon = \lambda\epsilon_0$ where $\lambda$ is the dimensionless noise scale factor. We realize different values of $\lambda$ by using gate-level unitary folding \cite{Schultz2022}. This method amplifies noise by scaling the entire circuit unitary, \( U \). For example, a noise amplification factor of \( \lambda = 3 \) is realized by replacing \( U \) with \( U U^\dagger U \), increasing the circuit depth proportionally. For \( \lambda = 2 \), one first considers a unitary \( U_h \) that corresponds to a subset of the qubit operations, having half the depth of the original circuit. Then, \( U \) is replaced by \( U U_h^\dagger U_h \), introducing some unknown approximation error in the noise amplification process. Importantly, the performance of NRE is known to be largely insensitive to imprecise noise amplification \cite{Amin2025}. This, in turn, relaxes the demanding requirement for precise noise amplification.

Figure~\ref{fig:TFIM_noisy_expts}(a) presents the measured ground-state energy as a function of noise scale factors. Due to inherent hardware noise, the measured energy is reduced by approximately $20\%$ from its ideal noiseless value at $\lambda=1$. In Fig.~\ref{fig:TFIM_noisy_expts}(b), we show the results after applying error mitigation with various different selections for noise scale factors. We find that NRE substantially improves the accuracy of noisy measurements of the QPU, yielding a final accuracy exceeding $96\%$, regardless of the specific noise amplification settings.
\newline
\section{Conclusion }\label{sec:conclusion}
We have demonstrated a QPU based on qubits coupled to a common computational resonator. We have calibrated and benchmarked the qubit-resonator CZ gate and the MOVE operation, which, in combination with single-qubit gates, form a universal set of quantum gates. Furthermore, we have generated genuine multipartite entanglement between six qubits in the form of a GHZ state with a readout-error mitigated fidelity of $0.86$ and provided a proof of principle demonstration on how to populate and calibrate higher level resonator states, eventually realizing a Jaynes-Cummings gate. Such a gate can be used as a building block for hybrid quantum computing schemes involving continuous and discrete variables \cite{liu2024}. In addition, we have shown a practical use case by employing our qubit-resonator QPU for QAOA applications including error mitigation. Finally, the presented proof of concept QPU shows a new paradigm for connectivity in superconducting architectures to match targeted applications. These include, for example, advanced quantum error correction strategies such as the color code \cite{Bombin2006, Landahl2011, Takada2024}. 

\medskip\noindent
\textbf{\large Acknowledgments}
\noindent
This work was supported by the German Federal Ministry of Education and Research through the projects DAQC (13N15686), QSolid
(13N16155) and Q-Exa (13N16065).

\medskip\noindent
\textbf{\large Data availability}
\noindent
The data that support the findings of this study are available from the corresponding author upon reasonable request.
\bibliography{Bibliography}

 \onecolumngrid



\newpage
\appendix

\setcounter{figure}{0}
\setcounter{equation}{0}
\setcounter{table}{0}
\counterwithout{equation}{section} 
\renewcommand{\thefigure}{S\arabic{figure}}
\renewcommand{\thetable}{S\arabic{table}}
\renewcommand{\theequation}{S\arabic{equation}}

\section{QPU Design and Fabrication}\label{sec:app_design}
\begin{figure}[h!]
    \centering
    \includegraphics[width=0.6\linewidth]{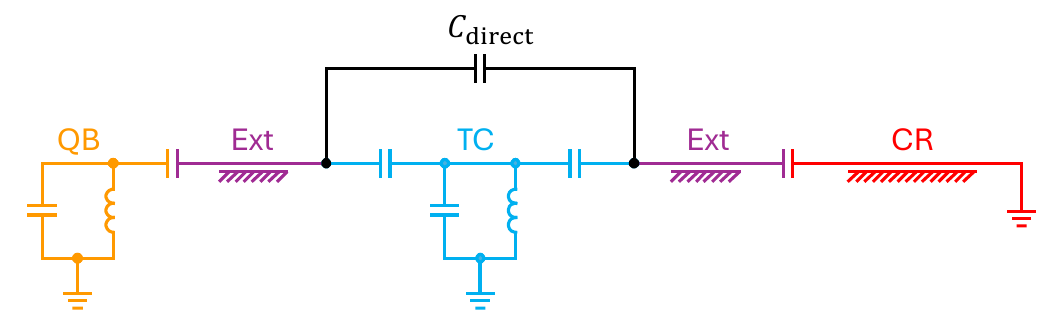}
    \caption{Simplified linearized circuit diagram of the QPU, showing a single qubit (QB) coupled via a tunable coupler (TC) to the computational resonator (CR). The couplings are mediated by coupler extender waveguides (Ext). The direct capacitance $C_\mathrm{direct}$ is part of of the TC geometry, and ensures that there is an idling configuration where the $ZZ$ interaction is zero. The remaining five qubits are connected to the CR in a similar fashion, each with a similar TC. In the model, each node also contains a parasitic capacitance to ground, which is omitted from the drawing for clarity.}
    \label{fig:app_design_schematic}
\end{figure}

The star topology QPU consists of a $\SI{4.2}{\milli\meter}$ long co-planar waveguide resonator in the quarter-wavelength configuration (computational resonator), six frequency-tunable couplers and six frequency-tunable qubits. The latter two are implemented using symmetric superconducting quantum interference devices (SQUIDs), enabling to tune the transition frequency close to zero \cite{Krantz2019}.
We model this QPU as a linearized circuit, where the SQUIDs have been replaced by linear inductors, as shown schematically for a single QB-TC-CR connection in Figure~\ref{fig:app_design_schematic}. Here, the localized qubit and coupler islands are modeled as pure capacitive structures (including a capacitance to ground at each island) extracted from finite element simulations of the respective geometry. The TC are coupled with interdigital capacitors near the open-circuit end of the CR. The QB-TC and TC-CR connections are formed by coupler extender waveguides, similar to those used in Ref.~\cite{Marxer2023}, which enable physically spacing out the qubits while still coupling them all near the maximum voltage of the computational resonator mode. The coupler extender waveguides and the computational resonator are modeled as transmission lines to accurately include their inductive contributions. All elements are referenced to a common ground plane in the design and the model.

We extract the mode frequencies in this model from the divergences of the frequency-dependent impedance $Z(\omega)$ calculated at the respective qubit or coupler island, or for the CR at its open end. We use the following procedure to numerically estimate the frequencies and coupling strengths of the full six-qubit star system. First, we tune all QB and TC modes far away from the resonator mode by adjusting their inductance values, and from the remaining model extract the bare resonator frequency $\omega_\mathrm{r}=4.3$~GHz. Then, we pairwise tune two modes on co-resonance, while tuning all remaining modes far away, and extract the corresponding normalized coupling strength $\beta_{ij} = g_{ij}/\omega$ from the frequency splitting $2g_{ij}$. For tuning the CR mode, we adjust the modelled transmission line length. Table~\ref{table:coupler_parameters} shows the modeled and experimentally measured coupling strengths.

\begin{table*}[h!]
\centering
\begin{tabular}{llrrrrrrr}
\hline
Parameter  & Description & QB\,1 & QB\,2 & QB\,3 & QB\,4 & QB\,5 & QB\,6 & \\
\hline
 $\beta_\mathrm{qc}$ &              & 0.0226 & 0.0226 & 0.0226 & 0.0226 & 0.0230 & 0.0230  \\
 $\beta_\mathrm{cr}$ & Design model & 0.0225 & 0.0225 & 0.0226 & 0.0226 & 0.0225 & 0.0225  \\
 $\beta_\mathrm{qr}$ &              & 0.0020 & 0.0020 & 0.0020 & 0.0020 & 0.0020 & 0.0020  \\
\hline
 $\beta_\mathrm{qc}$ & Experimentally measured & $0.017$ &  $0.022$   & $0.021$  &  $0.021$  &  $0.021$  & $0.022$  \\
\hline
\end{tabular}
\caption{\label{table:coupler_parameters} Coupler parameters extracted from the design model, and for QB-TC couplings measured experimentally. The subscripts q, c and r represent the QB, TC and CR, respectively.
}
\end{table*}

The quantum processing units (QPUs) were fabricated at the OtaNano Micronova cleanroom. A high-resistivity ($\rho$$>$10 k$\Omega$cm) n-type undoped (100) 6-inch silicon wafer, pre-cleaned to ensure a nonoxidized surface, served as the substrate. A 200-nm-thick high-purity niobium (Nb) layer was first deposited via sputtering to form the base superconducting circuit layer. Coplanar waveguides and capacitive structures were patterned using photolithography with a mask aligner, followed by reactive ion etching (RIE) to define the Nb features. Post-etching photoresist residuals were removed via ultrasonic cleaning in acetone and isopropanol (IPA), followed by nitrogen drying.
Josephson junctions for the transmon qubits were fabricated on individual dice via electron-beam lithography (EBL). A bilayer of methylmethacrylate/polymethylmethacrylate (MMA/PMMA) resist was employed, with development in a sequence of methyl isobutyl ketone (MIBK):IPA (1:3), methyl glycol, and IPA. Residual resist was removed via a oxygen plasma descum process.
The junctions were fabricated using electron-beam shadow evaporation technique, depositing two aluminum (Al) layers under ultrahigh vacuum. Lift-off in heated acetone finalized the junction structures. Airbridges for interlayer connections were formed by Al, followed by a second lift-off process.
Post-fabrication, the room-temperature resistance of the Josephson junctions was measured to verify junction integrity prior to cryogenic characterization. This process integrates high-yield lithography, contamination-minimized etching, and precise junction fabrication, ensuring robust superconducting circuits for quantum computing applications.

\section{Experimental Setup}\label{sec:exp_setup}
\begin{figure*}[h!]
    \centering
    \includegraphics[width=0.8\linewidth]{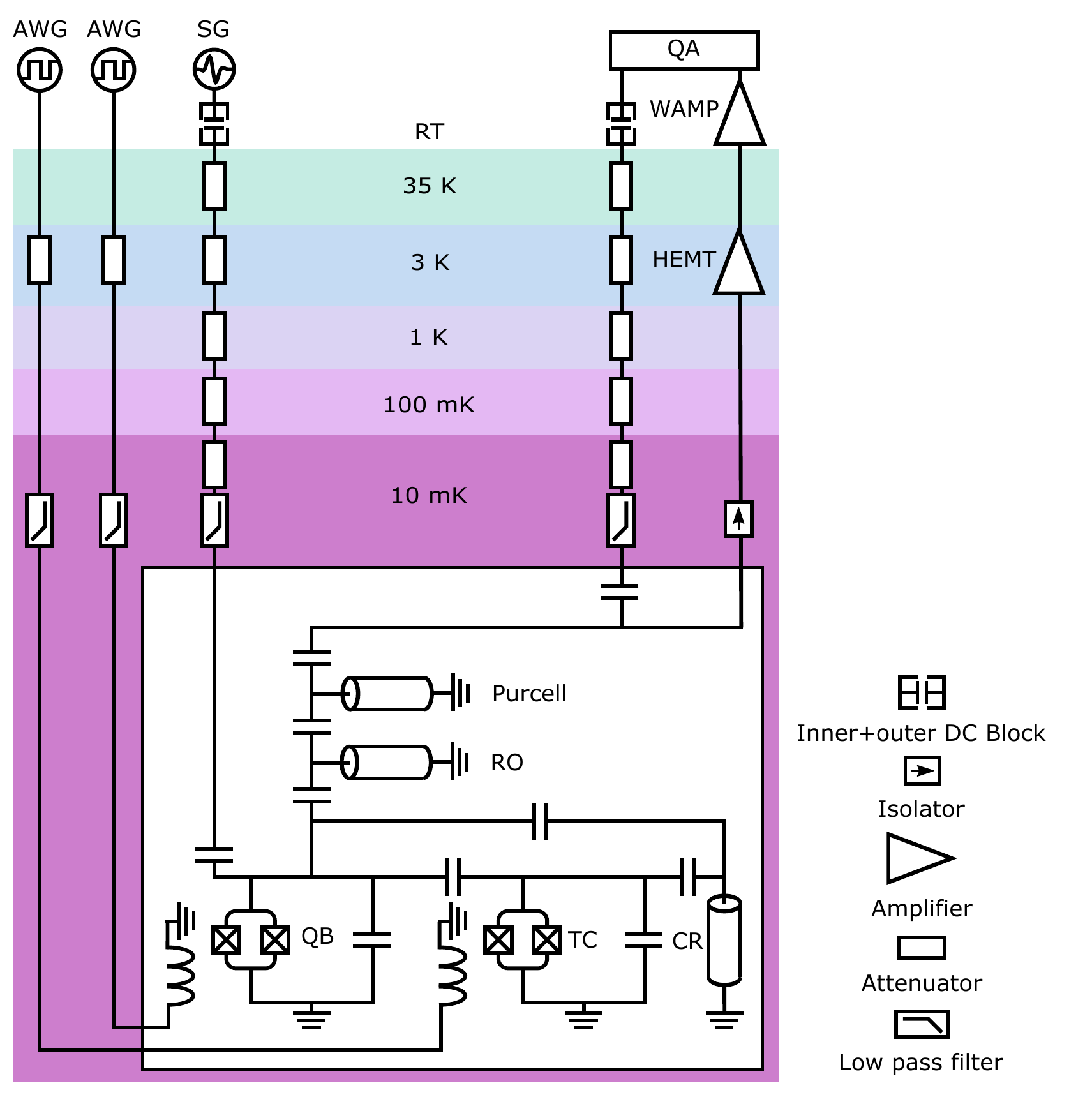}
    \caption{Experiment setup schematically depicting the QPU in its cryogenic environment. The QPU is connected via coaxial lines to control electronics at room temperature. The temperature of each stage is indicated.}
    \label{fig:setup}
\end{figure*}

Figure~\ref{fig:setup} illustrates the experimental setup with a simplified QPU circuit diagram featuring one qubit QB (including its readout resonator RO and Purcell filter), the computational resonator CR and a tunable coupler TC that couples the qubit to the computational resonator. The QPU and some electrical components are maintained at cryogenic temperature in a Bluefors XLD400 cryostat. The control electronics instruments, at room temperature, are connected to the QPU via coaxial cables. These contain microwave attenuators and filters distributed over the various temperature stages of the cryostat. Flux pulses and flux biases are generated by arbitrary waveform generators AWG (Zurich Instrument HDAWG8), the qubit drive pulses are generated by signal generators SG (Zurich Instrument SHFQC6+), and the readout pulses are generated and acquired by a quantum analyzer QA (contained in the same SHFQC6+ instrument). Before digitization, the readout pulses are amplified by a cryogenic amplifier based on a high electron mobility transistor HEMT and a warm amplifier WAMP. A group of isolators protects the readout resonator from noise.

\section{Analytic Description of Qubit--Resonator MOVE and CZ Operations}\label{sec:SWtrans}

We consider here the approximate diagonalization of a setup that can be modelled with the Hamiltonian 
\begin{equation}
\hat{H} = \hat{H}_0 + \hat{V},
\end{equation}
where \( \hat{H}_0 \) is an uncoupled (diagonal) Hamiltonian, and \( \hat{V} \) is a Hermitian operator describing the interactions between the eigenstates of $\hat H_0$. 
In particular, we concentrate on a system which comprises a qubit, a resonator, and a tunable coupler. Consequently, the unperturbed Hamiltonian 
can be written as
\begin{equation}
\hat{H}_0 = \hbar\omega_\mathrm{q} \hat{a}_\mathrm{q}^\dagger \hat{a}_\mathrm{q} + \frac{\hbar}{2} \alpha_\mathrm{q} \hat{a}_\mathrm{q}^\dagger \hat{a}_\mathrm{q}^\dagger \hat{a}_\mathrm{q} \hat{a}_\mathrm{q} + \hbar \omega_\mathrm{r} \hat{a}_\mathrm{r}^\dagger \hat{a}_\mathrm{r} +\hbar \omega_\mathrm{c} \hat{a}_\mathrm{c}^\dagger \hat{a}_\mathrm{c} + \frac{\hbar}{2} \alpha_\mathrm{c} \hat{a}_\mathrm{c}^\dagger \hat{a}_\mathrm{c}^\dagger \hat{a}_\mathrm{c} \hat{a}_\mathrm{c},
\end{equation}
where \( \omega_\mathrm{q} \), \( \omega_\mathrm{r} \) and \( \omega_\mathrm{c} \) are the (angular) frequencies of the qubit, resonator and coupler, respectively, $\alpha_{\rm q}$ 
is the anharmonicity of the qubit, \( \alpha_{\rm c} \) is the anharmonicity of the coupler, and \( \hat{a}_i \) and \( \hat{a}_i^\dagger\) are the annihilation and creation operators for the qubit, resonator and coupler, respectively, for $i\in\{\mathrm{q,r,c} \}$. The interaction part, $\hat{V} = \hat V_1 + \hat V_2$, describes the couplings between the qubit, resonator and coupler degrees of freedom. We have defined the interaction Hamiltonians as 
\begin{equation}
    \begin{split}
        \hat{V}_1 =& -\hbar g_\mathrm{qc}(\hat a_\mathrm{q}^{\dag} - \hat a_\mathrm{q})(\hat a_\mathrm{c}^{\dag} - \hat a_\mathrm{c}) -\hbar g_\mathrm{rc}(\hat a_\mathrm{r}^{\dag} - \hat a_\mathrm{r})(\hat a_\mathrm{c}^{\dag} - \hat a_\mathrm{c}),\\
        \hat V_2 =&-\hbar g_\mathrm{qr}(\hat a_\mathrm{q}^{\dag} - \hat a_\mathrm{q})(\hat a_\mathrm{r}^{\dag}-\hat a_\mathrm{r}).
    \end{split}
\end{equation}
Above, we denote the coupling strength between the qubit and the resonator with \( g_\mathrm{qr} \), the coupling strength between the qubit and the coupler with \( g_\mathrm{qc} \),  and the coupling strength between the resonator and the coupler with \( g_\mathrm{rc} \). 

Here, we assume that coupling to the coupler is dispersive, i.e., $g_\mathrm{qc} \ll |\Delta_\mathrm{qc}|$ and $g_\mathrm{rc} \ll |\Delta_\mathrm{rc}|$. However, we do not make such assumption for the qubit-resonator coupling $g_{\rm qr}$. Since the coupler is in the dispersive limit, we make a Schrieffer--Wolff transformation $\hat U_{\rm SW} = e^{\hat S}$ in order to eliminate the first-order interaction $\hat V_1$ with the coupler, and consequently obtain an effective Hamiltonian for the qubit-resonator system. In general, the Schrieffer--Wolff transformation can be expanded using the Baker--Campbell--Hausdorff lemma as
\begin{equation}
\hat U_{\rm SW} \hat H \hat U_{\rm SW}^{\dag} = \hat H + [\hat S,\hat H] + \frac{1}{2!}[\hat S,[\hat S,\hat H]] + \ldots.
\label{SW1}
\end{equation}
Here, we assume that $[\hat S,\hat H_0+\hat V_2] = - \hat V_1$, which removes the first-order dependence on $\hat V_1$ from the transformed Hamiltonian. 
The condition obtained above is fulfilled in our setup if we write $\hat S = \hat S_{\rm q} + \hat S_{\rm r}$, and define
~\cite{Yan_2018, heunisch2023tunable}
\begin{align}
    \hat{S}_\mathrm{q} &= \sum_{n_\mathrm{q},n_\mathrm{c} \in \{0,1\} } \sqrt{(n_\mathrm{q} + 1)(n_\mathrm{c} + 1)} \left[ \frac{g_\mathrm{qc}}{\Delta_\mathrm{q c} + n_\mathrm{q} \alpha_\mathrm{q} - n_\mathrm{c} \alpha_\mathrm{c} }\left( \hat{\pi}_\mathrm{q}^{n_\mathrm{q}+1,n_\mathrm{q}} \hat{\pi}_\mathrm{c}^{n_\mathrm{c},n_\mathrm{c}+1} - \hat{\pi}_\mathrm{q}^{n_\mathrm{q},n_\mathrm{q}+1} \hat{\pi}_\mathrm{c}^{n_\mathrm{c}+1,n_\mathrm{c}}\right)  \right.  \nonumber \\
    &- \left. \frac{g_\mathrm{q c}}{\Sigma_\mathrm{q c} + n_\mathrm{q} \alpha_\mathrm{q}  + n_\mathrm{c} \alpha_\mathrm{c}}\left( \hat{\pi}_\mathrm{q}^{n_\mathrm{q}+1,n_\mathrm{q}} \hat{\pi}_\mathrm{c}^{n_\mathrm{c}+1,n_\mathrm{c}} - \hat{\pi}_\mathrm{q}^{n_\mathrm{q},n_\mathrm{q}+1} \hat{\pi}_\mathrm{c}^{n_\mathrm{c},n_\mathrm{c}+1}\right)\right] , \\
    \hat{S}_r &= \sum_{n_\mathrm{r},n_\mathrm{c} \in \{0,1\} } \sqrt{(n_\mathrm{r} + 1)(n_\mathrm{c} + 1)} \left[ \frac{g_\mathrm{rc}}{\Delta_\mathrm{r c} - n_\mathrm{c} \alpha_\mathrm{c} }\left( \hat{\pi}_\mathrm{r}^{n_\mathrm{r}+1,n_\mathrm{r}} \hat{\pi}_\mathrm{c}^{n_\mathrm{c},n_\mathrm{c}+1} - \hat{\pi}_\mathrm{r}^{n_\mathrm{r},n_\mathrm{r}+1} \hat{\pi}_\mathrm{c}^{n_\mathrm{c}+1,n_\mathrm{c}}\right)  \right.  \nonumber \\
    &- \left. \frac{g_\mathrm{r c}}{\Sigma_\mathrm{r c} + n_\mathrm{c} \alpha_\mathrm{c}}\left( \hat{\pi}_\mathrm{r}^{n_\mathrm{r}+1,n_\mathrm{r}} \hat{\pi}_\mathrm{c}^{n_\mathrm{c}+1,n_\mathrm{c}} - \hat{\pi}_\mathrm{r}^{n_\mathrm{r},n_\mathrm{r}+1} \hat{\pi}_\mathrm{c}^{n_\mathrm{c},n_\mathrm{c}+1}\right)\right] , 
\end{align}
where the operators $\hat{\pi}_{k}^{n,m} = |n\rangle \langle m |$ act in the Hilbert space of $k\in \{ \mathrm{q,r,c}\}$, and we have defined $\Delta_{\mathrm{rc}} = \omega_\mathrm{r} - \omega_{\mathrm c}$, $\Delta_{\mathrm{qc}} = \omega_\mathrm{q} - \omega_{\mathrm c}$, $\Sigma_\mathrm{qc} = \omega_\mathrm{q} + \omega_{\mathrm c}$ and $\Sigma_\mathrm{rc} = \omega_\mathrm{r} + \omega_{\mathrm c}$. 
Consequently, the transformed Hamiltonian can be written in second order in the coupling strengths as
\begin{equation}
\hat{U}_{\mathrm{SW}}\hat{H}\hat{U}^\dagger_{\mathrm{SW}} = \hat{H}_0 +\hat{V}_2 + \frac{1}{2} [\hat{S}, \hat{V}_1].
\label{SW2}
\end{equation}
In our architecture, we typically consider that the coupling strength $g_\mathrm{q r}$ is almost an order of magnitude smaller than $g_\mathrm{q c}$ and $g_\mathrm{rc}$.

After performing the Schrieffer--Wolff transformation (evaluating Eq.~\eqref{SW2}) \cite{Yan_2018,sung2021}, we obtain corrections to qubit and resonator frequencies and effective couplings between the relevant (coupler-dressed) states that govern the MOVE and CZ operations. 
By additionally setting the energy of the ground state to zero, we obtain that the qubit and resonator frequencies and the effective couplings between the relevant single- and two-excitation eigenstates can be written as
\begin{align}
    \tilde{\omega}_\mathrm{q} &= \omega_\mathrm{q}  + \frac{g_\mathrm{q c}^2}{\Delta_\mathrm{q c}} - \frac{2g_\mathrm{q c}^2}{\Sigma_\mathrm{q c} + \alpha_\mathrm{q} } + \frac{g_\mathrm{q c}^2}{\Sigma_\mathrm{q c}},\label{eq:perturbative_hamiltonian_parameters_omega}\\
    \tilde{\omega}_\mathrm{r} &= \omega_\mathrm{r}  + \frac{g_\mathrm{r c}^2}{\Delta_\mathrm{q c}}  - \frac{g_\mathrm{r c}^2}{\Sigma_\mathrm{r c}},\label{eq:perturbative_hamiltonian_parameters_omega}\\
    \tilde{g}_{\mathrm{MOVE}} &\equiv \tilde{g}_{0e,1g} = g_\mathrm{q r} + \frac{ g_\mathrm{q c}  g_\mathrm{rc} }{2} \left(\frac{1}{\Delta_\mathrm{q c}} + \frac{1}{\Delta_\mathrm{r c}} 
    -  \frac{1}{\Sigma_\mathrm{q c}} - \frac{1}{\Sigma_\mathrm{r c}} \right), \label{effective_coupling_gtilde}\\
    \tilde{g}_{1e,2g} &= \sqrt{2} \tilde{g}_{0e,1g}, \label{eq:1e2g}\\
    \tilde{g}_{\mathrm{CZ}} &\equiv  \tilde{g}_{1e,0f} = \sqrt{2}\left[ g_\mathrm{q r} + \frac{ g_\mathrm{q c}  g_\mathrm{r c} }{2} \left(\frac{1}{\Delta_\mathrm{r c}} + \frac{1}{\Delta_\mathrm{q c} + \alpha_\mathrm{q}} 
    -  \frac{1}{\Sigma_\mathrm{r c}} - \frac{1}{\Sigma_\mathrm{q c} + \alpha_\mathrm{q}} \right) \right]. \label{eq:1e0f}
\end{align}
Here, we have defined that $\tilde{g}_{0e,1g}$ is the coupling strength between the (coupler-dressed) states $|eg0\rangle$ and $|gg1\rangle$, $\tilde{g}_{1e,2g}$ between the states $|eg1\rangle$ and $|gg2\rangle$, and $\tilde{g}_{1e,0f}$ between the states $|eg1\rangle$ and $|fg0\rangle$ (state labelling goes as $|\rm qubit,coupler,resonator\rangle$).
We assume here that the coupler is in the ground state. Furthermore, we also assume that the qubit and coupler frequencies are constant and, thus, 
the relations obtained above are a good approximation only for idling gates or for longer (adiabatic) gates dominated by a long interaction time. Our effective Hamiltonian (not shown) also includes other non-zero terms that couple states, such as $|gg0\rangle$ to $|eg1\rangle$ and $|gg0\rangle$ to $|gg2\rangle$, but these leakage transitions can be neglected under the rotating-wave approximation. 
The effective coupling strengths $\tilde{g}_{0e,1g}$ and $\tilde{g}_{1e,0f}$ are 
responsible for the MOVE and CZ gate dynamics, and from hereafter we refer to them as $\tilde{g}_{\mathrm{MOVE}}$  and  $\tilde{g}_\mathrm{{CZ}}$, respectively. We note that Eq.~\eqref{eq:1e2g} indicates that populating the $|eg1\rangle$ state during the MOVE gate results in undesired population transfer to the $|gg2\rangle$ state which is outside of the computational subspace. Furthermore, Eq.~\eqref{eq:1e2g} and Eq.~\eqref{eq:1e0f} can be further generalised. For $n$ photons in the resonator, the effective coupling strengths can be written as $\tilde{g}_{(n-1)e,ng} = \sqrt{n} \tilde{g}_{0e,1g}$ and $\tilde{g}_{n e,(n-1)f} = \sqrt{n} \tilde{g}_{1e,0f}$. As is the case in the Jaynes-Cummings physics, coupling strengths are seen to scale with the photon number as $\sqrt{n}$. 

Utilising Eq.~\eqref{eq:1e0f} and truncating the Hamiltonian given in Eq.~\eqref{SW2} to a 5-level subspace spanned by $\{|gg 0\rangle,|eg0\rangle,|gg1\rangle,|eg1\rangle,|fg0\rangle \}$, we obtain analytical expressions for the population of the $|eg1\rangle$ state and the acquired conditional phase during the CZ gate operation. In doing so, we assume that the leakage outside of the truncated subspace is negligible, and that at the CZ gate operational point, i.e., at $\Tilde{\omega}_\mathrm{q} = \Tilde{\omega}_\mathrm{r} - \alpha_q$, the single-excitation swapping $\tilde{g}_{\mathrm{MOVE}}$ can be neglected because $|\tilde{\omega}_\mathrm{q} -\tilde{\omega}_\mathrm{r}|\gg \tilde{g}_{\mathrm{MOVE}}$. We obtain the effective time-evolution operator as $\hat U = \exp(-\textrm{i}\tilde{H}t)$, where $\tilde{H}$ is the truncated Hamiltonian in the coupler-dressed eigenbasis 
and, by applying it to the initial $|eg1\rangle$ state, we obtain that the population in the state $|eg1\rangle$ can be written as~\cite{papivc2024charge}
\begin{equation}\label{eq:perturbative_P11}
    P_{|eg1\rangle}(t) = 1 - \frac{2 \tilde{g}_{\mathrm{CZ}}^2}{\tilde{\Omega}^2}\left[ 1 - \cos(\tilde{\Omega} t)\right],
\end{equation}
where $\tilde{\Omega} = \sqrt{(\tilde{\Delta} - \alpha_\mathrm{q})^2 + 4\tilde{g}_{\mathrm{CZ}}^2 }$ and $\tilde{\Delta} =  \tilde{\omega}_\mathrm{r} - \tilde{\omega}_\mathrm{q}$. Eq.~\eqref{eq:perturbative_P11} suggests that during the CZ gate, the population periodically returns to the $|eg1\rangle$ state exhibiting the Rabi physics~\cite{scully1997quantum}.
The conditional phase that the state $|eg1\rangle$ picks up during the CZ gate can also be analytically extracted and is given as~\cite{papivc2024charge}
\begin{align}\label{eq:perturbative_cphase}
    \phi(t) &= \frac{1}{2}\left[ (\alpha_\mathrm{q} - \tilde{\Delta})t + \pi \left(1 - \mathrm{sign}\{\cos(\tilde{\Omega} t/2) \} \right) \right] +  \mathrm{arctan}\left( \frac{\tilde{\Delta} - \alpha_\mathrm{q}}{\tilde{\Omega}} \tan(\tilde{\Omega} t/2)\right),
\end{align}
where the first term composes a geometric phase, while the latter is related to the dynamic phase. In the experiment, the gate time $\tau_\mathrm{{CZ}}$ is fixed, and $\tilde{\Delta}$ and $\tilde{\Omega}$ are tuned via the amplitudes of the qubit and coupler flux pulses such that $\phi(\tau_{\mathrm{{CZ}}}) = \pi$.

\begin{figure*}
	\begin{center}
\includegraphics[width=0.55\linewidth,angle=0,clip]{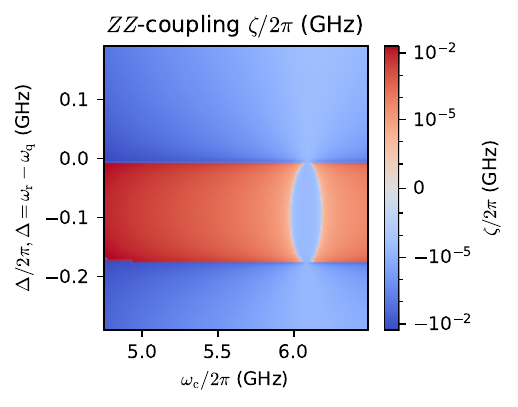}
    \caption{Longitudinal $ZZ$-coupling landscape as a function of the bare coupler frequency, $\omega_c$, and the bare qubit-resonator detuning, $\Delta$.}
    \label{fig:zz_landscape}
    \end{center}
\end{figure*}

The analytical expression for the conditional phase is an approximation valid near the resonance condition, $\Tilde{\omega}_\mathrm{q} \approx \Tilde{\omega}_\mathrm{r} - \alpha_\mathrm{q}$, and describes the conditional phase acquired when applying a non-adiabatic CPhase gate. To complement the discussion of the conditional phase, we present a numerical simulation of the longitudinal $ZZ$-coupling strength between a qubit and a computational resonator, defined as 
\begin{equation}
    \zeta = \tilde{\omega}_{eg1} + \tilde{\omega}_{gg0} - \tilde{\omega}_{gg1} - \tilde{\omega}_{eg0}.
\end{equation}
Including the lowest three energy levels and assuming dimensionless coupling strengths $\beta_\mathrm{qr} = 0.00197$, $\beta_\mathrm{qc} = 0.0219$ and $\beta_\mathrm{cr} = 0.02264$, a resonator frequency $\omega_\mathrm{r}/(2\pi) = \SI{4.3}{\giga \hertz}$, a qubit anharmonicity $\alpha_\mathrm{q}/(2\pi) = \SI{-0.187}{\giga \hertz}$ and a coupler anharmonicity $\alpha_\mathrm{c}/(2\pi) = \SI{-0.11}{\giga \hertz}$, we simulate the $ZZ$-coupling for a broad range of the bare qubit-resonator detuning, $\Delta = \omega_\mathrm{r} - \omega_\mathrm{q}$, and the bare coupler frequency, $\omega_\mathrm{c}$. 
In contrast to the $ZZ$-coupling between two qubits, the results shown in Fig.~\ref{fig:zz_landscape}, exhibit distinct behavior depending on the sign of the qubit-resonator detuning due to the vanishing anharmonicity of the resonator. The simulation results demonstrate that $\zeta$ can reach several tens of MHz, and also reveal the existence of an idling configuration, where the $ZZ$-coupling vanishes (see the white ellipse-shaped contour). We note that the value of $\zeta$ determines the speed limit for a fully adiabatic gate, while in this work we have employed the non-adiabatic gate protocol.

\begin{figure*}
	\begin{center}
\includegraphics[width=0.7\linewidth,angle=0,clip]{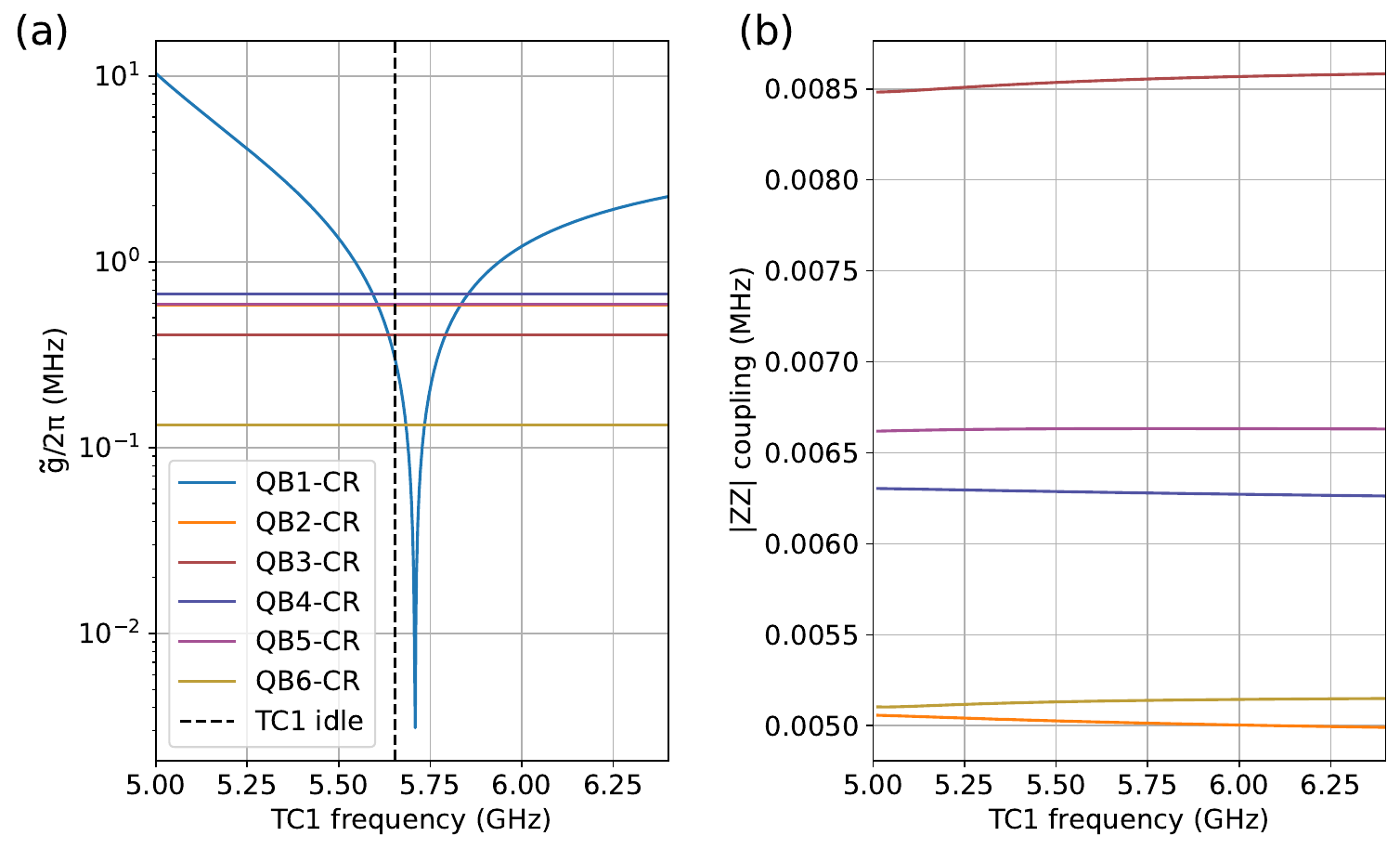}
    \caption{(a) Calculated effective transverse coupling $\tilde{g}$ between the CR and the spectator qubits during a CZ gate involving QB1 and the CR. (b) Corresponding simulated $ZZ$-coupling between the CR and the spectator qubits.}
    \label{fig:spectator_xx_zz_coupling}
    \end{center}
\end{figure*}

Next, we investigate the residual transverse and longitudinal couplings between the computational resonator and the spectator qubits during qubit-resonator gate operation. These interactions are important to consider because they impact spectator coherence times and spectator leakage. We perform a numerical simulation using the chip-specific characteristics and vary the frequency of the tunable coupler implementing the qubit-resonator gate. In Fig.\,\ref{fig:spectator_xx_zz_coupling}, we show the theory results on the residual couplings to the five spectator qubits for a CZ gate between QB1 and the computational resonator.

For the simulation of the $ZZ$-coupling, the computational states are chosen from the eigenstates of the idling Hamiltonian with the largest overlap with the uncoupled qubit states. We calculate these eigenstates from the system Hamiltonian with six qubits, six tunable couplers and the resonator. The idling configuration is chosen such that the overall $ZZ$-coupling is minimized. During the CZ gate QB1 is taken to a frequency where its second excited state is in resonance with the state where both QB1 and the resonator are excited. The coupling between QB1 and the resonator is then achieved by sweeping the corresponding tunable coupler closer to the qubit. The resulting $ZZ$-couplings between the resonator and the spectators are obtained by calculating the eigenvalues of the same Hamiltonian as a function of the TC1 frequency. 
We determine the effective transverse coupling $\tilde{g}$ [cf. Eq.\,\eqref{effective_coupling_gtilde}] between the central resonator and the qubits in a coupler-hybridized basis. We plot the resulting transverse coupling strength for the spectator qubits together with the coupling strength between QB1 and the resonator in Fig.\,\ref{fig:spectator_xx_zz_coupling}(a). Figure\,\ref{fig:spectator_xx_zz_coupling}(b) shows the $ZZ$-coupling between the spectator qubits and the resonator. We find that the transverse coupling between the resonator and the target qubit (QB1) strongly depends on the coupler frequency, as desired for a tunable coupling scheme. In contrast, the transverse coupling to the spectators is below $\SI{1}{\mega \hertz}$. Similarly, the spectator $ZZ$-coupling remains almost constant and is of the order of a few kHz.

\section{Calibration and Benchmarking}\label{sec:cal_bench_appendix}
In the following, we describe the calibration procedure of the qubit-resonator MOVE operation and the CZ gate, and report on the basic characterization of the QPU.

\subsection{Calibration of the MOVE Operation}\label{Sec:MOVE_fine_VZ}
The initial calibration of the MOVE operation is presented in the main text, therefore only the fine calibration of the operating point and the calibration of the single-qubit phase correction of the MOVE operation are discussed here.
\begin{figure*}
	\begin{center}
 \includegraphics[width=\linewidth,angle=0,clip]{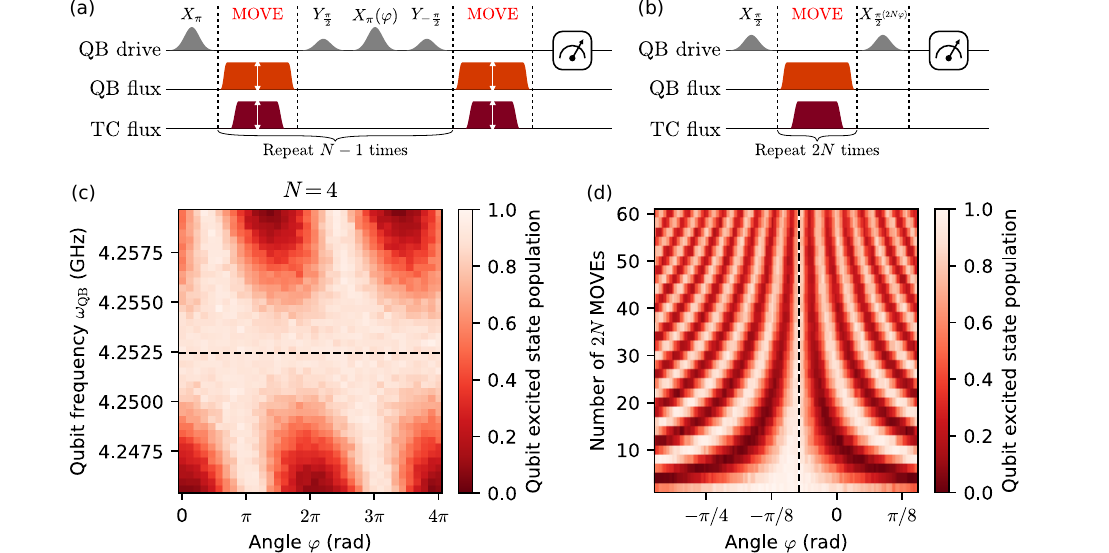}
    \caption{Calibration of the MOVE operation. (a), (b) Pulse schedules used for the fine calibration of the MOVE operating point and the calibration of the single-qubit phase correction of the MOVE, respectively. The schedules are described in detail in the supplementary text. (c) Experimental data showing the fine calibration measurement to optimise the flux pulse amplitude of the qubit $Z$-pulse, i.e., the diabatic qubit frequency $\omega_\mathrm{QB}$ at the operation point of the MOVE. The optimal qubit frequency is visualized by the horizontal dashed line. (d) Experimental data for the calibration of the VZ phase correction. The optimal phase is visualized by the vertical dashed line.}
    \label{fig:MOVE_fine_VZ}
    \end{center}
\end{figure*}
The fine calibration experiment for the MOVE operation individually calibrates both the qubit and coupler flux pulse amplitudes by making the MOVE insensitive to the relative phase of qubit and resonator, which signals a complete population transfer.
As shown in the pulse schedule in Fig.~\ref{fig:MOVE_fine_VZ}(a), we first excite the qubit using an $X_\pi$ gate and subsequently move this excitation to the computational resonator. We adjust the relative phase between qubit and resonator by applying a physical $Z$-rotation to the qubit by an angle $\varphi$. This is achieved by applying the following pulse sequence to the qubit: a $\pi/2$-pulse around the $y$-axis, followed by a rotation around the $x$-axis by an angle $\varphi$, and another $\pi/2$-pulse around the $y$-axis but rotating in the opposite direction. We repeat the sequence consisting of the MOVE operation and the $Z$-rotation $N-1$ times, and apply a final MOVE operation before measuring the state of the qubit. The total number of moves, $N$, should be chosen to be an even number, such that most of the population is back in the qubit at the end of the experiment. If the MOVE operation does not fully transfer the state, the excited state probability measured at the end of the sequence oscillates as a function of the relative phase $\varphi$. This oscillation is due to the quantum interference of the residual superposition state of the qubit after the first MOVE with the part of the state that has been moved to the computational resonator and back to the qubit. The more MOVE operations are applied in the fine calibration sequence, the more sensitive the resulting interference pattern is to a deviation from the optimal operating point.
Experimental data are depicted in Fig.~\ref{fig:MOVE_fine_VZ}(c) for a total number of $N=4$ MOVE operations. For the data shown, we vary the amplitude of the qubit flux pulse and convert this voltage into the qubit frequency using the measured qubit flux dispersion. We take the average along the relative phase $\varphi$, and approximate the resulting data by a second order polynomial to obtain the optimal qubit frequency where the average qubit excited state probability is maximal. We perform a similar measurement (not shown here) to optimize the amplitude of the coupler flux pulse for fixed qubit flux pulse amplitude.

Next, we determine the optimal virtual Z (VZ) phase for the MOVE qubit, which corrects for the single-qubit phase that is accumulated depending on the frequency trajectory during the MOVE operation. Since we cannot directly probe the phase of a state in the computational resonator, we apply the MOVE operation in pairs and thus calibrate the common phase \cite{bao2022fluxonium}, i.e., the phase obtained when moving an excitation from the qubit to the resonator and back. To achieve this, we consider the pulse schedule shown in Fig.~\ref{fig:MOVE_fine_VZ}(b). First, an equal superposition state is prepared in the qubit by applying an $X_\frac{\pi}{2}$-pulse. Next, we move this superposition state back and forth between the qubit and the resonator, amplifying the accumulated single-qubit phase. Finally, we apply another $X_\frac{\pi}{2}$-pulse with the rotation axis defined by $\varphi$ in order to map the azimuthal angle of the qubit state to an excited state probability. If the phase of the qubit state was not changed by the application of the MOVE operations, we expect an excited state probability of 1, independent of $N$. In Fig.~\ref{fig:MOVE_fine_VZ}(d), we show experimental data, where we vary the rotation axis of the second $\pi/2$-pulse, $\varphi$, for an even number of MOVE operations. We average the data along the number of MOVE operations to extract the VZ phase correction. For demonstration purposes, we show the phase landscape for up to 60 MOVE operations, while for the benchmarking data we typically average over $2N = 2, 4, ..., 10$ MOVE operations only.

\subsection{Calibration of the CZ Gate}\label{Sec:CZ_fine_VZ}
The initial calibration of the CZ gate is presented in the main text, therefore only the fine optimization of the operating point, including the measurement of the conditional phase and the population exchange, as well as the calibration of the single-qubit phase corrections of the CZ gate are discussed here.
\begin{figure*}
\begin{center}
\includegraphics[width=\linewidth,angle=0,clip]{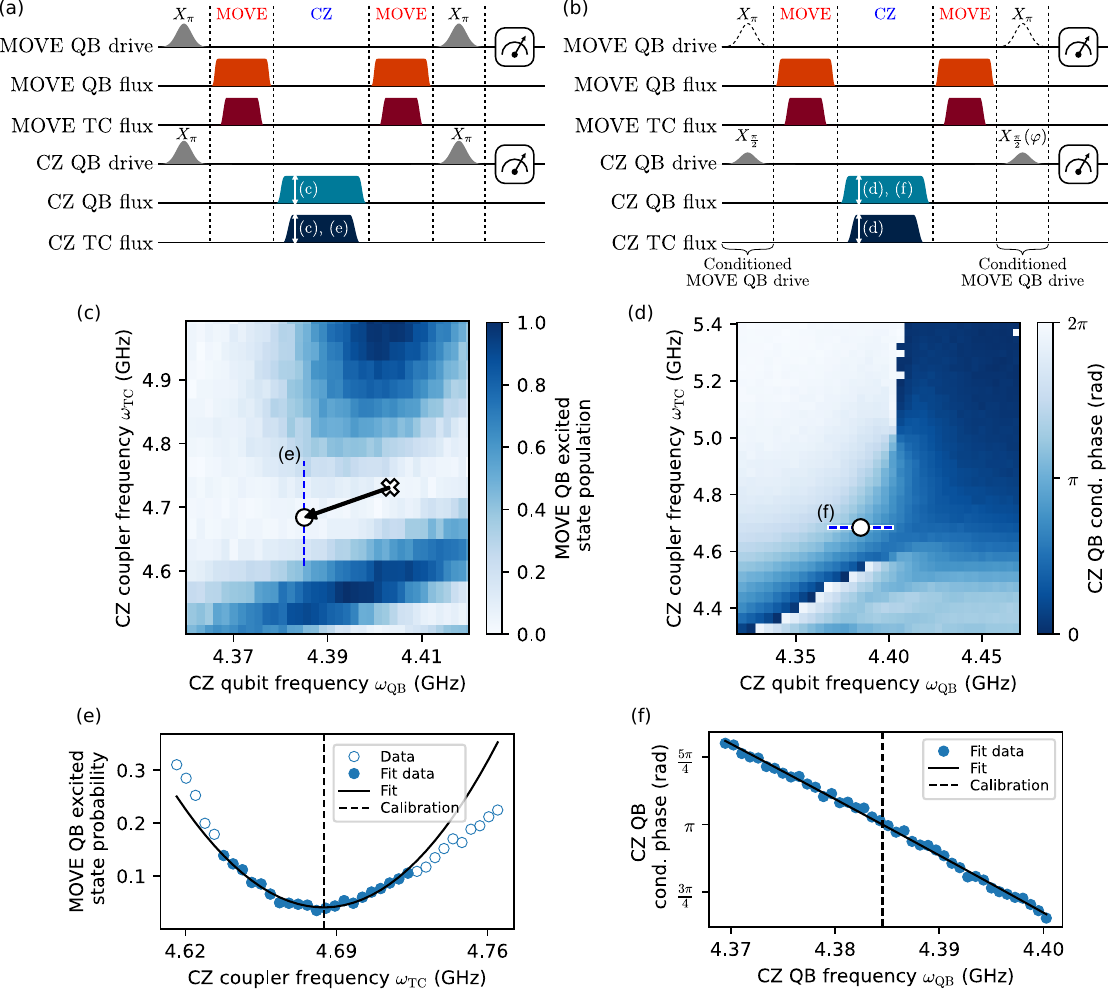}
    \caption{Calibration of the CZ gate. (a), (b) Pulse schedules to calibrate the population exchange and conditional phase, respectively. The flux pulses implementing the CZ gate to be calibrated are annotated with the measurement data figures, for which the respective pulse amplitude is varied. (c), (d) Experimental data showing the population exchange and conditional phase measured as a function of both the qubit and coupler frequencies realized during the gate operation by varying the respective flux pulse amplitudes. The blue dashed lines indicate the sweep ranges used for the one-dimensional data shown in (e) and (f), respectively. The white markers represent the operating point of the CZ gate inferred from the initial calibration discussed in the main text (cross) and the final calibration after several iterations of optimizing the flux pulse parameters based on the measurement of the population exchange and the conditional phase (circle). (e) Experimental data of the excited state probability of the MOVE qubit versus the coupler frequency during the gate. We approximate the minimum in the excited state probability with a quadratic fit (black solid line) that is applied to a subset of the data points (blue solid circles) obtained by removing data points from either side based on the R-squared value of the fit. (f) Experimental data of the conditional phase versus the qubit frequency during the gate. The optimal operating point of the CZ gate is indicated by the black dashed lines in (e) and (f).}
    \label{fig:CZ_fine}
    \end{center}
\end{figure*}
We refine the calibration of the CZ gate by iteratively optimizing the qubit and coupler flux pulse parameters based on the measurement of the conditional phase and the population exchange, respectively. Typically, three cycles of iteration are sufficient to converge to the optimal operating point. We start by optimizing the coupler frequency during the gate using the sequence shown in Fig.~\ref{fig:CZ_fine}(a). To prepare the $\ket{eg1}$ state, we excite both the MOVE and CZ qubit with a $X_\pi$ pulse and transfer the state of the MOVE qubit to the computational resonator. Next, we initiate the $\ket{eg1} \rightarrow \ket{fg0}\rightarrow \ket{eg1}$ population oscillations by applying flux pulses to the CZ qubit and the coupler connecting the CZ qubit and the computational resonator. Finally, we move the population of the computational resonator back to the MOVE qubit and apply $X_\pi$ pulses to the MOVE and CZ qubit to de-excite them to the ground state, assuming the CZ gate performed a full population oscillation. We extract the operating point of the CZ gate from the minimum in the excited state probability of the MOVE qubit. The excited state probability of the CZ qubit also reveals the population oscillations, however, in general we obtain higher signal-to-noise for the MOVE qubit, as the readout is optimized to distinguish the qubit states $\ket{g}$ and $\ket{e}$, whereas the CZ qubit oscillates between $\ket{g}$ and $\ket{f}$. When varying the flux pulse amplitudes of both the CZ qubit and coupler pulses, we obtain the two-dimensional population landscape shown in Fig.~\ref{fig:CZ_fine}(c). We use the qubit and coupler flux dispersions to express the flux pulse amplitudes in terms of the respective component frequency during the gate. We obtain the optimal coupler frequency by fitting the excited state probability of the MOVE qubit for a one-dimensional subset of the data as shown in Fig.~\ref{fig:CZ_fine}(e).

Next, we optimize the qubit frequency during the CZ gate based on the measurement of the conditional phase. The pulse schedule used for measuring the phase accumulation of the CZ qubit depending on the state of the computational resonator is shown in Fig.~\ref{fig:CZ_fine}(b). 
The CZ qubit is prepared in a superposition state by applying an $X_\frac{\pi}{2}$-pulse to enable the measurement of the phase accumulation. The state of the computational resonator during the CZ gate depends on whether we apply the initial $X_\pi$-pulse on the MOVE qubit before moving the qubit state to the resonator. After the first MOVE operation, we apply flux pulses to the CZ qubit and the coupler connecting the CZ qubit and the computational resonator to implement the CZ gate. Finally, we apply an additional $\frac{\pi}{2}$-pulse on the CZ qubit with rotation axis $\varphi$ to map the azimuthal angle of the qubit state to the $Z$-basis, i.e., to an excited state probability. From the measurement of the excited state probability versus the angle $\varphi$, we determine the single-qubit phase accumulation of the CZ qubit for the two cases where the resonator is in the ground state or the first excited state. Figure~\ref{fig:CZ_fine}(d) shows the measurement of the conditional phase when sweeping the frequencies during gate operation of both the CZ qubit and its tunable coupler. We obtain the optimal qubit frequency during the CZ gate from a fit to a one-dimensional subset of the data with fixed coupler frequency at the point where the conditional phase is equal to $\pi$ as shown in Fig.~\ref{fig:CZ_fine}(f).

\begin{figure*}
	\begin{center}
\includegraphics[width=\linewidth,angle=0,clip]{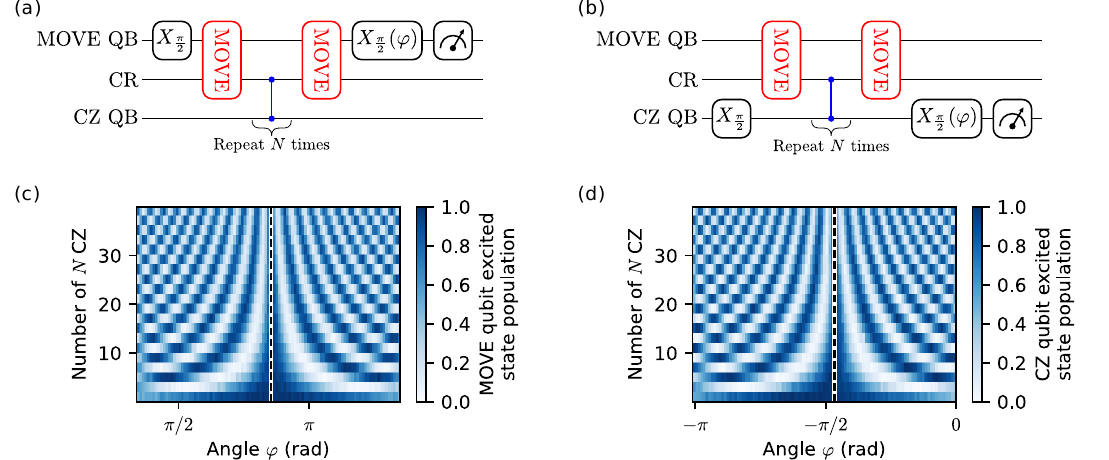}
    \caption{Calibration of the single-qubit phase corrections for the CZ gate. (a), (b) Pulse schedules to extract the single-qubit phase corrections of the computational resonator and the CZ qubit, respectively. (c), (d) Experiment data of the excited state probability of the qubit with the Ramsey sequence corresponding to the circuits shown in (a) and (b). We vary the angle $\varphi$ and the total number of CZ gates $N$ to obtain the optimal phase corrections. Since the additional contributions ($\mathrm{VZ}_\mathrm{MOVE}$, $\Delta\cdot t_\mathrm{CZ}$) have been accounted for during the execution of the experiment, the vertical black dashed lines directly represent the optimal VZ rotation angle.}
    \label{fig:CZ_VZ}
    \end{center}
\end{figure*}

Next, we discuss the calibration of the single-qubit phase corrections of the CZ gate for both the computational resonator and the CZ qubit using the circuits shown in Fig.~\ref{fig:CZ_VZ}(a) and (b), respectively. First, we prepare an equal superposition state in the target qubit by applying an $X_\frac{\pi}{2}$-pulse. The target qubit can be either the MOVE qubit or the CZ qubit. 
Before applying the CZ gate $N$ times between the computational resonator and the CZ qubit, we move the population of the MOVE qubit to the resonator. Finally, we apply another MOVE operation followed by a $X_\frac{\pi}{2}$-pulse in order to map the azimuthal angle of the qubit state to an excited state probability. If the phase of the target qubit has not been changed by applying the CZ gates, we expect the excited state probability to be equal to 1. 
We note that we apply the MOVE operations in the circuit shown in Fig.~\ref{fig:CZ_VZ}(b) even though there is no population in the MOVE qubit, because the flux pulses that implement a MOVE operation between the MOVE qubit and the  computational resonator can affect the single-qubit phase accumulation of the CZ qubit due to flux cross-talk.
Experiment data of the calibration of the single-qubit phase corrections for the CZ gate are shown in Fig.~\ref{fig:CZ_VZ}(c) and (d). If the CZ qubit is the target qubit, the phase accumulation that we measure originates purely from the frequency change of the CZ qubit during the CZ gate (ignoring flux cross-talk).  However, if the MOVE qubit is the target qubit, the total accumulated phase between the $\frac{\pi}{2}$-pulses generally contains different contributions. To measure the pure single-qubit phase accumulation of the computational resonator due to the dynamic frequency tuning during the CZ gate, we ensure that the previously calibrated VZ correction of the MOVE operation, $\mathrm{VZ}_\mathrm{MOVE}$, is applied. Furthermore, we account for the time-dependent phase due to the relative rotation of the rotating frames of the MOVE qubit and the computational resonator by adding a phase, $\Delta \cdot N \cdot t_\mathrm{CZ} = (\omega_\mathrm{QB} - \omega_\mathrm{CR}) \cdot N \cdot t_\mathrm{CZ}$, which is proportional to the detuning of the MOVE qubit and the resonator, as well as to the total time in between the MOVE operations, $N \cdot t_\mathrm{CZ}$.

\subsection{Characterization of the Computational Resonator}
\label{sec:app_cr_char}

\begin{figure*}
	\begin{center}
\includegraphics[width=\linewidth,angle=0,clip]{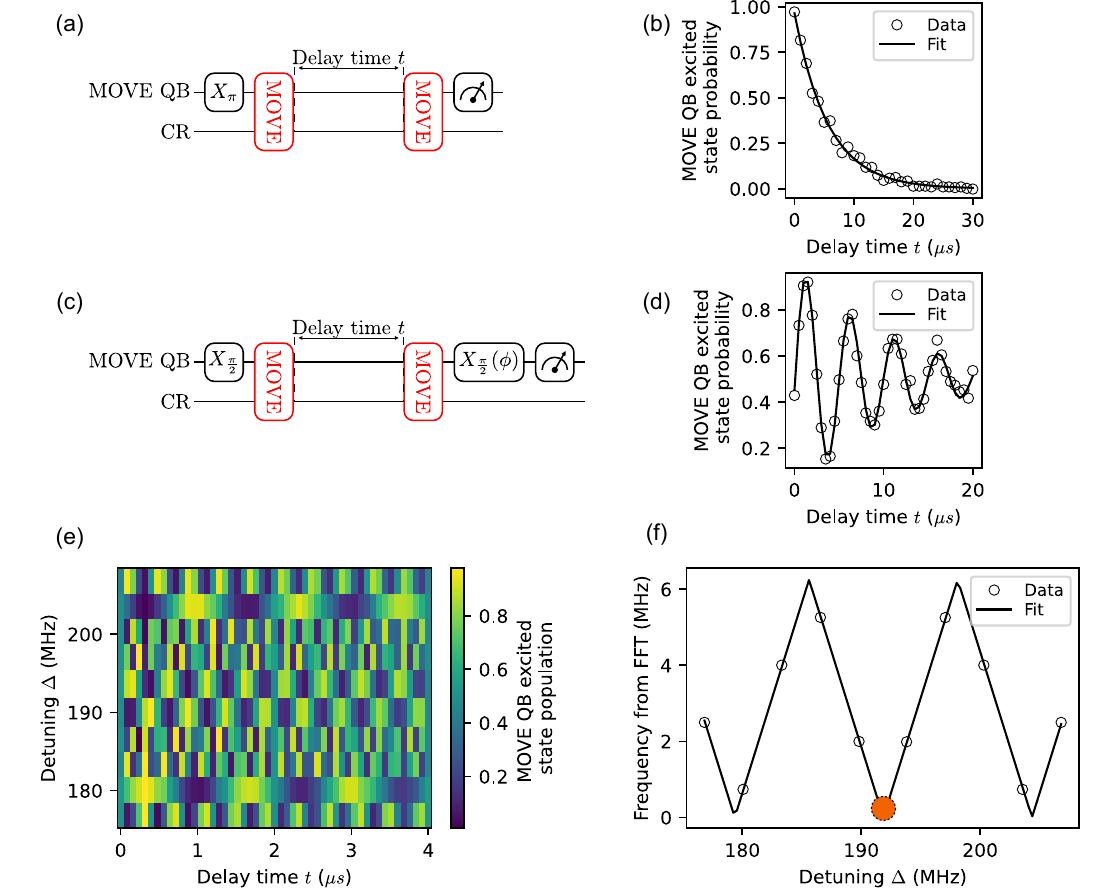}
    \caption{Characterization of the computational resonator. (a), (b) Circuit and corresponding experiment data to determine the relaxation time of the computational resonator $T_1$. (c), (d) Circuit and measurement data from which the dephasing time of the resonator $T^*_2$ is determined. (e) Experiment data with additional sweep over the virtual detuning $\Delta$ for the circuit shown in (c). (f) Fast Fourier transform of the qubit excited state probability versus delay time $t$ shown in (e). We fit a sawtooth model to obtain the value of the virtual detuning corresponding to the frequency difference of the qubit and the computational resonator. The sweep range of the detuning is centered around the initial estimate for the qubit-resonator detuning obtained from a similar measurement but with much larger Nyquist frequency of approximately $250\,\mathrm{MHz}$.  Therefore, the qubit-resonator detuning (indicated by the orange circle) is obtained as the value of the detuning $\Delta$ that is closest to the center of
    the sweep range and where the Nyquist fit is equal to zero.}
    \label{fig:CR_t_f}
    \end{center}
\end{figure*}

We characterize the frequency and coherence properties of the computational resonator by employing the MOVE operation to (de)populate the resonator. The circuit to measure the relaxation time is shown in Fig.~\ref{fig:CR_t_f}(a): we excite one of the qubits using an $X_{\pi}$-pulse. After moving the excitation to the resonator, it will decay according to the resonator relaxation rate. After a variable delay time, we use a second MOVE operation to move the state back and measure the remaining population in the qubit. The data in Fig.~\ref{fig:CR_t_f}(b) shows the exponential decay from which we extract the resonator relaxation time $T_1 = 5.53 \pm \SI{0.32}{\micro \second}$.
Several factors contribute to the small resonator relaxation time. 
Simulations suggest that the close proximity of two qubit flux lines to the grounded end of the computational resonator results in strong inductive coupling and hence Purcell decay, limiting the resonator $T_1$ time to $\SI{14.5}{\micro \second}$. Multiple other Purcell decay channels from other flux and drive lines may further reduce $T_1$. The coupling between the resonator and the control lines is mostly mediated by the high-dielectric constant of the silicon substrate. Switching to a flip-chip architecture, where the control lines can be on a different substrate, can strongly reduce this loss channel. Furthermore, optimizing the coplanar waveguide geometry by adjusting the width of the centre conductor and the gap to the grounding plane is expected to reduce the energy participation ratio (EPR). In addition, the design of the tunable couplers can be improved to further decrease the EPR. In this work, waveguide extenders are incorporated in the coupling structure with large interdigitated capacitors on each side of the tunable coupler. Removing the extender between the computational resonator and the tunable coupler may further reduce the EPR. By addressing these design and fabrication aspects, we are confident
that improving the resonator relaxation time by an order of magnitude is possible.

To measure the $T^*_2$ coherence time of the resonator, we employ Ramsey interferometry as shown in Fig.~\ref{fig:CR_t_f}(c). This measurement is based on a technique outlined in Ref.~\cite{chu2017}. We prepare a superposition state in the MOVE qubit by applying an $X_\frac{\pi}{2}$-pulse and use the MOVE operation to transfer the prepared state into the resonator. After a certain delay time $t$, we use a subsequent MOVE operation to move the superposition back to the qubit. Finally, we apply an $X_\frac{\pi}{2}(\phi)$-pulse with a rotation axis that forms an angle of $\phi$ with the X-axis. The excited state probability of the qubit at the end of the circuit oscillates at the difference frequency of the qubit and the computational resonator, which can be quite large, i.e., on the order of $100\,\mathrm{MHz}$. We use a virtual detuning $\Delta$ to advance the phase of the second $\pi/2$ pulse by $\phi= \Delta t$ to effectively slow down the oscillations in the excited state probability of the qubit, such that the observed oscillation frequency $\omega = \omega_\mathrm{QB} - \omega_\mathrm{CR} - \Delta$ is comparable to the inverse of the $T^*_2$ time. Both the oscillation frequency $\omega$ and the dephasing time $T^*_2$, are extracted from an exponentially decaying sinusoidal fit. The measurement data in Fig.~\ref{fig:CR_t_f}(d) shows the Ramsey fringes from which we extract $T^*_2 = 10.9 \pm 1.0 \mu$s. By additionally sweeping the virtual detuning as shown in Fig.~\ref{fig:CR_t_f}(e), we can measure the eigenfrequency of the computational resonator. The excited state probability of the qubit versus the delay time is Fourier transformed via FFT. In Fig.~\ref{fig:CR_t_f}(f), we show the extracted dominant frequency component versus the virtual detuning. We fit a sawtooth model to extract the value of the detuning, which corresponds to the detuning of the qubit and the computational resonator, i.e., $\Delta = \omega_\mathrm{QB} - \omega_\mathrm{CR}$.

\subsection{QPU Characterization}\label{sec:qpu_charact_appendix}
We summarize the QPU characterization and single-qubit benchmarking data in Tab.~\ref{table:QPU_parameters}. The data for the coherence times, readout fidelity, and single-qubit gate fidelity has been taken over a time frame of $238$ hours. The statistics consist of $208$ data points for the coherence times, $210$ data points for the readout fidelity and $49$ data points for the single-qubit gate fidelities, which have been characterized individually and simultaneously on all qubits. In between the individual benchmarking measurements, the QPU has been recalibrated by an automatic service. The Josephson energy $E_\mathrm{J}$ has been determined by fitting the coupler flux dispersion. The energy ratio $E_\mathrm{J}/E_\mathrm{C}$ is determined by using the design parameters for the charging energy $E_\mathrm{C}$. The individual gate fidelities for the MOVE operation and for the CZ gate are shown in Tab.~\ref{table:2qb_gate_parameters}. The error bars are determined from the corresponding fit errors by error propagation. We implement single-qubit gates using the derivative removal by adiabatic gate (DRAG) method with a cosine shaped envelope of the in-phase component \cite{motzoi2009simple}. For the MOVE operation and for the CZ gate, we use cosine flux pulses on the qubits and square pulses smoothed by truncated Gaussian ramps for the tunable couplers. The respective gate durations, including buffer time to account for residual flux pulse distortions affecting the performance of the qubit-resonator operations, are summarized in Tab.~\ref{table:gate_times}.
\begin{table*}[h!]
\centering
\begin{tabular}{llccccccc}
\hline
Parameter  & Description & QB\,1 & QB\,2 & QB\,3 & QB\,4       & QB\,5 & QB\,6 & CR \\
\hline
 $f$\,(GHz)  &  Transition frequency & $4.67$ & $4.47$ & $4.41$ & $4.52$ & $4.63$ & $4.93$ & $4.22$  \\
 $T_1$\,(µs) & Lifetime          & $25.6 \pm 4.1$     &  $44.0 \pm 5.3$   & $55.3 \pm1.1$  &  $46.2 \pm9.0$  &  $45.9 \pm8.6$  & $30.6 \pm5.6$  &  $5.53 \pm0.32$ \\
 $T_2^*$\,(µs) & Dephasing time & $36.5\pm6.1$  & $27.1\pm3.4$ & $29.0\pm4.3$ & $22.2\pm2.0$ & $51.5\pm9.9$ & $31.9\pm8.2$ & $10.9\pm1.0$
 \\
 $T_2^\mathrm{e}$\,(µs) &  Hahn echo dephasing time & $44.2\pm7.1$ & $56.0\pm5.5$ & $42.3\pm4.5$ & $29.9\pm2.2$ & $58.6\pm8.8$ & $43.8\pm8.1$ &   \\
 $T_\mathrm{q}$\,(mK) &  Effective qubit temperature  & $46.3$ & $42.0$ & $43.6$ & $40.9$   & $43.0$ & $45.8$ &   \\
 $E_\mathrm{J}/h$\,(GHz) &  Josephson energy   & $14.8$  & $13.8$ & $13.3$ & $13.9$   & $14.6$ & $16.5$ &   \\
 $E_\mathrm{J}/E_\mathrm{C}$ &  Energy ratio   & $74.2$  &  $68.8$  & $66.5$  &  $69.6$  & $73.0$  & $82.3$ &   \\
 $F_\mathrm{RO}\,(\%)$  &  Simultaneous readout fidelity   & $98.3\pm0.3$ & $98.6\pm0.2$ & $98.7\pm0.3$ & $99.1\pm0.2$ & $98.9\pm0.2$ & $98.7\pm0.6$ &   \\
 $F_\mathrm{sq,ind}\,(\%)$  &  Individual SQG fidelity  & $99.93\pm0.02$  & $99.94\pm0.04$ & $99.96\pm0.02$ & $99.96\pm0.01$ & $99.96\pm0.01$ & $99.89\pm0.3$ &   \\
 $F_\mathrm{sq,sim}\,(\%)$  &  Simultaneous SQG fidelity  & $99.93\pm0.02$ & $99.92\pm0.04$ & $99.96\pm0.03$ & $99.95\pm0.01$ & $99.59\pm0.01$ & $99.87\pm0.3$ &   \\
\hline
\end{tabular}
\caption{QPU performance. If provided, the uncertainty corresponds to the empirical standard deviation measured over a $\SI{238}{\hour}$ period. For the qubit/resonator frequency, the standard deviation is on the order of $\SI{10}{\kilo \hertz}$.}\label{table:QPU_parameters}
\end{table*}

\begin{table*}[h!]
\centering
\begin{tabular}{llcccccc}
\hline
Parameter  & Description & QB\,1 & QB\,2 & QB\,3 & QB\,4 & QB\,5 & QB\,6  \\
\hline
 $F_\mathrm{mm}\,(\%)$  &  Double MOVE fidelity & $99.11\pm0.05$ & $99.34\pm0.03$ & $99.00\pm0.03$ & $99.30\pm0.03$ & $98.31\pm0.06$ & $97.95\pm0.10$   \\
 $F_\mathrm{cz}\,(\%)$  &  CZ fidelity   & $98.90\pm0.05$ & $98.75\pm0.04$ & $98.97\pm0.03$ & $98.04\pm0.08$ & $98.53\pm0.14$ & $96.61\pm0.05$  \\
\hline
\end{tabular}
\caption{Individual double MOVE and CZ fidelities.}\label{table:2qb_gate_parameters}
\end{table*}

\begin{table*}[h!]
\centering
\begin{tabular}{llccccccc}
\hline
Parameter  & Description & QB\,1 & QB\,2 & QB\,3 & QB\,4 & QB\,5 & QB\,6  \\
\hline
 $\tau_\mathrm{s}$\,(ns) & Single-qubit gate duration & $40$ &  $40$   & $40$  &  $40$  &  $40$  & $40$ \\
 $\tau_\mathrm{ro}$\,(ns) & Readout pulse duration & $800$ &  $800$   & $800$  &  $800$  &  $800$  & $800$ \\
 $\tau_\mathrm{m}$\,(ns) & MOVE operation duration & $88$ &  $80$   & $96$  &  $80$  &  $96$  & $96$ \\
 $\tau_\mathrm{cz}$\,(ns) & CZ gate duration & $96$ &  $80$   & $80$  &  $112$  &  $96$  & $80$ \\
\hline
\end{tabular}
\caption{Duration of single-qubit gates, multiplexed readout and qubit-resonator operations.
}\label{table:gate_times}
\end{table*}

\section{Coherence Limit}\label{sec:coherence_limit}
To estimate the coherence limit $F^\mathrm{c}_\mathrm{cz}$ for the MOVE operation and for the CZ gate, we employ the weak dissipation result $F^\mathrm{c}$, valid at absolute zero, from Ref.~\cite{Abad2022}
\begin{equation}\label{Eq:general_coherence_limit}
F^\mathrm{c} = 1 - \frac{d \tau}{2(d+1)}\sum_{k=1}^N \left(\gamma_1^{(k)} + \gamma_\varphi^{(k)}\right).
\end{equation}
Here, $N$ is the number of involved components, $d$ is the Hilbert space dimension, $\tau$ the gate duration, and $\gamma_1^{k}$ $(\gamma_\varphi^{(k)})$ is the energy relaxation (dephasing) rate for component $k$. Equation~\eqref{Eq:general_coherence_limit} is valid as long as we do not leave the computational subspace during the operation. Furthermore, Eq.~\eqref{Eq:general_coherence_limit} contains the implicit assumption that dephasing is caused by a white noise spectrum. In case of additional low frequency noise with $1/f$ noise spectral density, Eq.~\eqref{Eq:general_coherence_limit} would contain an additional quadratic contribution for the dephasing rate. In the parking configuration, we verify that we are dominated by white noise since the envelope of the Ramsey oscillations is exponential and not Gaussian. We thus only consider the linear white noise contributions, although the contribution of Gaussian noise is expected to be significant during the gate operation due to the flux pulse, detuning the qubit away from the flux-insensitive parking configuration. In addition, we also neglect hybridization effects during the gate, thus, we obtain an upper bound for the coherence limit.

To apply Eq.~\eqref{Eq:general_coherence_limit} for a single MOVE operation with a duration $\tau_\mathrm{m}$, we treat it as an effective single-qubit gate and 
simplify the calculation by assuming that the decoherence is caused only by the qubit for the first half of the gate, and only by the computational resonator for the second half, thus
\begin{equation}\notag
F_\mathrm{m}^\mathrm{c} = 1 - \frac{\tau_\mathrm{m}}{6}\left(\gamma_1^\mathrm{q} + \gamma_1^\mathrm{r} + \gamma_\varphi^\mathrm{q} + \gamma_\varphi^\mathrm{r}\right),
\end{equation}
where the subscript q (r) refers to the MOVE qubit (computational resonator). The coherence limit of the double MOVE operation satisfies $F_\mathrm{mm}^\mathrm{c} = (F_\mathrm{m}^\mathrm{c})^2$. During the CZ gate, we leave the computational space and apply the formula from Ref.~\cite{Abad2024}
\begin{equation}
F_\mathrm{cz}^\mathrm{c}= 1 - \frac{1}{2}\tau_\mathrm{cz}\gamma_1^\mathrm{q} - \frac{3}{10}\tau_\mathrm{cz} \gamma_1^\mathrm{r} - \frac{61}{80}\tau_\mathrm{cz}\gamma_\varphi^\mathrm{q} - \frac{29}{80}\tau_\mathrm{cz}\gamma_\varphi^\mathrm{r},
\end{equation}
where $\tau_\mathrm{cz}$ is the CZ gate time. In the presence of a thermal qubit (resonator) population of $n_\mathrm{q}$ ($n_\mathrm{r}$) photons, the coherence limits reduce further according to
\begin{equation}
F_\mathrm{m}^\mathrm{c} \to F_\mathrm{m}^\mathrm{c} - \frac{1}{3}\tau_\mathrm{m} \gamma_1^\mathrm{q} n_\mathrm{q} - \frac{1}{3}\tau_\mathrm{m} \gamma_1^\mathrm{r} n_\mathrm{r},
\qquad 
F_\mathrm{cz}^\mathrm{c} \to F_\mathrm{cz}^\mathrm{c} - \frac{9}{5}\tau_\mathrm{cz} \gamma_1^\mathrm{q} n_\mathrm{q} - \frac{7}{5}\tau_\mathrm{cz} \gamma_1^\mathrm{r} n_\mathrm{r}.
\end{equation}
The qubit and resonator decay rates are given in Tab.~\ref{table:QPU_parameters}, and the gate durations of both MOVE and CZ in Tab.~\ref{table:gate_times}.  We neglect thermal effects since we operate the QPU in the saturation regime of Planck statistics, i.e., for qubit temperatures below $\SI{50}{\milli \kelvin}$ \cite{Mariantoni2010, Gandorfer2025}.

We can also estimate the effect of decoherence on the GHZ state preparation. For this, we consider the transpiled circuit we employ for $N$-qubit GHZ state generation shown in Fig.~\ref{fig:ghz_bench}(a), which can be schematically written as
\begin{equation}
\underbrace{Y_{\frac{\pi}{2}}}_{(1)} + \underbrace{\mathrm{MOVE}}_{(2)} + \underbrace{\mathrm{CZ}}_{(3)} + \underbrace{\mathrm{MOVE}}_{(4)} + \underbrace{Y_{\frac{\pi}{2}} + X_\pi}_{(5)}.
\end{equation}
Step $(1)$ corresponds to a $\pi/2$-rotation about the $y$-axis which is applied to all $N$ qubits. In step $(2)$, we employ a MOVE operation between a dedicated MOVE qubit and the computational resonator. Step $(3)$ corresponds to a cascade of $(N-1)$ CZ operations, followed by a MOVE operation in $(4)$ which transfers the state of the resonator back to the MOVE qubit. Finally, we apply a $\pi/2$-rotation about the $y$-axis and a $\pi$-rotation about the $x$-axis to each of the CZ qubits. In case $F_i$ corresponds to the coherence-limited fidelity of step $(i)$, we approximate the coherence limit of the entire circuit by assuming 
\begin{equation}\label{Eq:GHZ_fidelity_limit_ansatz}
F_\mathrm{GHZ} = F_1 F_2 F_3 F_4 F_5.
\end{equation}
In the following, we assume that qubit $l$ is used as the MOVE qubit. The individual fidelity contributions are estimated by
\begin{equation}\label{Eq:GHZ_coherence_limit_contributions}
F_1 = \prod_k F_\mathrm{s}^{(k)}, \qquad F_2 = F_4 = F_\mathrm{m}^{(l)} \prod_{k \neq l} F_\mathrm{i, m}^{(k)}, \qquad F_3 = \prod_{k \neq l} F_\mathrm{cz}^{(k)} \left(F_\mathrm{i,cz}^{(k)}\right)^{N-2}, \qquad F_5 = \prod_{k \neq l} (F_\mathrm{s}^{(k)})^2 (F_{\mathrm{i,s}}^{(l)})^2.
\end{equation}
Here, $F_\mathrm{s}^{(k)}$ denotes the single-qubit gate fidelity coherence limit for qubit $k$, given by \cite{Abad2022}
\begin{equation}\label{Eq:SQG coherence limit}
F_\mathrm{s}^{(k)} = 1 - \frac{\tau_\mathrm{s}}{3}\left(\gamma_1^{(k)} + \gamma_\varphi^{(k)}\right) - \frac{2}{3}n_\mathrm{q}^{(k)}\gamma_1^{(k)} \tau_\mathrm{s}.
\end{equation}
Here, $F_\mathrm{m}^{(l)}$ corresponds to the MOVE operation coherence limit and $F_\mathrm{i, m}^{(k)}$ corresponds to the fidelity limit for qubit $k$ which idles during the MOVE operation. The quantity $F_\mathrm{cz}^{(k)}$ corresponds to the fidelity limit for the CZ gate between the computational resonator and qubit $k$ and $F_\mathrm{i,cz}^{(k)}$ denotes the idling fidelity of qubit $k$ during the CZ gates with other qubits. The exponent $N-2$ in $F_3$ results from the fact that there is one CZ gate per qubit applied and each qubit idles during the remaining $(N-2)$ CZ gates. Finally, $F_{\mathrm{i,s}}^{(l)}$ is the idling fidelity limit for the MOVE qubit during the final two single-qubit gates. For simplicity, we dropped the superscript c for the coherence limits. We exploit the fact that the coherence limit is in first order independent of the applied gate. Therefore, we approximate the idling fidelities by the measured single-qubit gate fidelities. Thus, $F_\mathrm{i,s} = F_\mathrm{s}$ and $F_{\mathrm{i,s}}^{(l)}$ as well as $F_\mathrm{i,cz}^{(k)}$ are treated like an effective single-qubit gate with the same gate time as the respective qubit-resonator operation. As a consequence, Eq.~\eqref{Eq:GHZ_fidelity_limit_ansatz} can be expressed as
\begin{equation}\label{Eq:GHZ_coherence_limit_estimation}
F_\mathrm{GHZ} = (F_\mathrm{m}^{(l)})^2 F_\mathrm{s}^{(l)}\left(F_{\mathrm{i, s}}^{(l)}\right)^2 \prod_{k \neq l} \left(F_\mathrm{s}^{(k)} \right)^3  \left(F_\mathrm{i, m}^{(k)}\right)^2 \left(F_\mathrm{i, cz}^{(k)}\right)^{N-2} F_\mathrm{cz}^{(k)}.
\end{equation}

As a consistency check, we employ Eq.\,\eqref{Eq:GHZ_coherence_limit_estimation} for the measured simultaneous single-qubit gate fidelities from Tab.\,\ref{table:QPU_parameters} and the two-qubit gate fidelities from Tab.\,\ref{table:2qb_gate_parameters}. Idling infidelities are modelled via Eq.\,\eqref{Eq:SQG coherence limit} using the QPU parameters from Tab.\,\ref{table:QPU_parameters} and the gate durations from Tab.\,\ref{table:gate_times}. In our particular case, we have employed QB3 as the MOVE qubit while the CZ cascade has been applied to all remaining qubits. We estimate $F_\mathrm{GHZ} \simeq 0.85$ which is close to the measured GHZ fidelity of 0.86, thereby demonstrating consistency of our model. Next, we use Eq.\,\eqref{Eq:SQG coherence limit} to estimate the coherence limit of the GHZ fidelity. As for the coherence limit estimation for the individual gates, hybridization effects with tunable couplers are neglected. However, we take the finite temperature into account this time, since even for a qubit temperature of $\SI{40}{\milli \kelvin}$, finite temperature effects accumulate due to the large amount of gates in the circuit. Since we cannot directly measure the temperature of the computational resonator, we perform a work case estimation and assume its temperature to coincide with the hottest qubit in Tab.~\ref{table:QPU_parameters}, $T_\mathrm{r} = \SI{46.3}{\milli \kelvin}$. For the six qubit GHZ state, we then find a fidelity limit of $F_\mathrm{GHZ} \simeq 0.910$. After multiplying with the readout fidelities, this limit reduces to $0.842$. As a comparison, we measured a GHZ fidelity of $0.815$ without REM and $0.860$ with REM. Although the model Eq.~\eqref{Eq:GHZ_coherence_limit_estimation} is relatively simple, we conclude that the main effect of GHZ infidelity results from decoherence. 

We particularly observe that the resonator life time is significantly shorter than the life time of the qubits. Since the resonator is involved in all qubit-resonator operations, this low life time is expected to set a dominant limitation on the GHZ state fidelity. For that, we employ the approximation that the qubits and the computational resonator are $T_1$ limited and drop all dephasing terms (for the resonator, this is a very accurate assumption). Simultaneously, we assume that the qubit-resonator operations add significantly more infidelity than the single-qubit gates, implying that we treat all single-qubit gates as ideal. We furthermore assume a temperature of absolute zero and, for simplicity, equal duration $\tau$ of CZ gate and MOVE operation, implying $F_{\mathrm{i,m}}^{(k)} = F_{\mathrm{i,cz}}^{(k)}$. With these rough approximations, we have
\begin{equation}
F_\mathrm{GHZ} \simeq \prod_{k \neq l} \left(F_\mathrm{i, cz}^{(k)}\right)^N F_\mathrm{cz}^{(k)} = \prod_{k \neq l} \left(1 - \frac{\tau}{3}\gamma_1^{(k)}\right)^N \left(1 - \frac{\tau}{2}\gamma_1^{(k)} - \frac{3}{10}\gamma_1^\mathrm{r}\right).
\end{equation}
In the limit $N \gg 1$, $\gamma_1^{(k)} \tau \ll 1$, $\gamma_1^{\mathrm{r}} \tau \ll 1$, we can further approximate by
\begin{equation}
F_\mathrm{GHZ} \simeq \prod_{k \neq l} e^{-N\frac{\tau}{3}\gamma_1^{(k)}} \prod_{k \neq l} e^{-\frac{\tau}{2}\gamma_1^{(k)} - \frac{3\tau}{10}\gamma_1^\mathrm{r}} = e^{-\tau \left(\frac{N}{3} + \frac{1}{2} \right) \sum_{k \neq l}\gamma_1^{(k)}} e^{-(N-1)\frac{3\tau}{10}\gamma_1^\mathrm{r}},
\end{equation}
where we use an exponential approximation, analogous as has been used for the derivation of the Porter-Thomas distribution in Ref.~\cite{Boixo2018}. With this approximation, $F_\mathrm{GHZ}$ decouples into a qubit and a resonator decay term
\begin{equation}\label{Eq:GHZ_limit_exp_decay}
F_\mathrm{GHZ} \simeq \underbrace{e^{-\tau\left(\frac{N}{3} + \frac{1}{2}\right)(N-1)\bar{\gamma}_1^\mathrm{q}}}_{\mathrm{Qubit\,decay}} \cdot \underbrace{e^{-(N-1)\frac{3\tau}{10}\gamma_1^\mathrm{r}}}_{\mathrm{Resonator\,decay}},
\end{equation}
where $\bar{\gamma}_1^\mathrm{q}$ is the mean energy decay rate of the involved CZ qubits. By comparing the decay constants, the constraint that the resonator life time limits the GHZ fidelity is then given by 
\begin{equation}\label{Eq:practical_CR_dominance_condition}
\gamma_1^\mathrm{r} \gg \left(\frac{10N}{9} + \frac{5}{3}\right) \bar{\gamma}_1^\mathrm{q}.
\end{equation}
For our six qubit QPU, this implies $\gamma_1^\mathrm{r} \gg 8.3\bar{\gamma}_1^\mathrm{q}$. For our QPU, we have $\gamma_1^\mathrm{r} \simeq 6.7 \bar{\gamma}_1^\mathrm{q}$, implying that this condition is not met. As such, the decoherence of the computational resonator is expected to have a significant effect on the GHZ coherence limit, but the dominating error contribution is the depolarization of the qubits during idling gates. In fact, setting $\gamma_1^\mathrm{r} = 0$ in Eq.~\eqref{Eq:GHZ_coherence_limit_estimation} raises $F_\mathrm{GHZ}$ from $0.842$ to $0.867$ (taking finite readout fidelity into account). Equation~\eqref{Eq:practical_CR_dominance_condition} also implies that for $N\gg1$, it can be beneficial to use the qubit with the lowest life time as MOVE qubit since during the long CZ cascade, it always idles in the ground state and does not contribute to $\bar{\gamma}_1^\mathrm{q}$.

\section{Jaynes-Cummings gate}\label{JC_gate_appendix}
\subsection{Time evolution in the rotating frame}
Let us consider here the dynamics of a quantum state $|\psi\rangle$, governed by the time-dependent Hamiltonian $\hat{H}(t) = \hat{H}_0 + \hat{V}(t)$. Here, we have separated the full Hamiltonian into a time-independent part $\hat H_0$, 
and into an interaction part $\hat{V}(t)$ which can be time dependent. In general, $\hat{H}_0$ and $\hat{V}(t)$ do not commute.

We define a change in the rotating frame by 
a unitary operator, and here we choose $\hat{R}(t) = e^{i{\hat{H}_0}t/\hbar}$, which transforms the state vector as follows $\ket{\psi}' = \hat{R}(t) \ket{\psi}$.    
The time evolution in the rotating frame is given by the (modified) Schrödinger equation:
\begin{equation}
        i\hbar \partial_t \ket{\psi}' = \underbrace{\hat{R}(t) \hat{V} \hat{R}^\dagger(t)}_{\hat{H}'}\ket{\psi}'.\label{eq:hamiltonian_rotatin_frame}
\end{equation}
The unitary time evolution operators in the two frame are related according to
\begin{equation}
    \hat{U}'(t,t_0) = \hat{R}(t) \hat{U}(t,t_0) \hat{R}^\dagger(t_0) \label{eq:time_evolution_operator_transformation}.
\end{equation}
\subsection{Derivation of the time evolution of the Jaynes-Cummings gate}
The Rabi Hamiltonian describes the time evolution of a two-level system, i.e., a qubit, interacting with a bosonic mode, here called a computational resonator. In the lab frame and ignoring the zero-point energy of the computational resonator, the Hamiltonian is given by
\begin{equation}
        \hat{H} = \hat{H}_\mathrm{resonator} + \hat{H}_\mathrm{qubit} + \hat{H}_\mathrm{interaction}
        = \hbar\omega_\mathrm{r}\hat{a}^\dagger\hat{a} + \hbar\omega_\mathrm{q}\frac{\pauliz}{2} + \frac{\hbar\Omega}{2}(\hat{\sigma}^+ +\hat{\sigma}^-)(\hat{a} + \hat{a}^\dagger), 
\end{equation}
Using Eq.~\eqref{eq:hamiltonian_rotatin_frame} we can transform the Hamiltonian from the Schrödinger picture (lab frame) to the interaction picture (computational frame), which is defined by the choice of rotating frame,  $\hat{H}_\mathrm{0} = \hat{H}_\mathrm{resonator} + \hat{H}_\mathrm{qubit}$
\begin{equation}
    \hat{H}_\mathrm{comp} = 
    \frac{\hbar\Omega}{2} e^{i\hat{H}_\mathrm{0}t/\hbar} (\hat{a}\hat{\sigma}^- + \hat{a}^\dagger\hat{\sigma}^+ + \hat{a}\hat{\sigma}^+ + \hat{a}^\dagger\hat{\sigma}^-) e^{-i\hat{H}_\mathrm{0}t/\hbar}. \\
\end{equation}
Using the Baker-Campbell-Hausdorff lemma ($e^XYe^{-X} = Y + [X, Y] + 1/2! [X,[X,Y]]+ ... + 1/n![X,[X,...[X,Y]...]]$)
and the commutation relations $[\hat{a}^\dagger\hat{a},\hat{a}] = -\hat{a}$, $[\hat{a}^\dagger\hat{a},\hat{a}^\dagger] = \hat{a}^\dagger$, and $[\pauliz, \hat{\sigma}^\pm] = \pm2\hat{\sigma}^\pm$, we have
\begin{equation}
     \hat{H}_\mathrm{comp} = \frac{\hbar\Omega}{2}\left( \hat{a}\hat{\sigma}^-e^{-i(\omega_\mathrm{r} + \omega_\mathrm{q})t} +\hat{a}^\dagger\hat{\sigma}^+e^{i(\omega_\mathrm{r} + \omega_\mathrm{q})t}+\hat{a}\hat{\sigma}^+e^{-i(\omega_\mathrm{r} -\omega_\mathrm{q})t}+\hat{a}^\dagger\hat{\sigma}^-e^{i(\omega_\mathrm{r} - \omega_\mathrm{q})t}\right).
\end{equation}
According to the rotating wave approximation (RWA), the two quickly oscillating terms may be ignored. In this case, we end up with the Jaynes-Cummings Hamiltonian in the interaction picture:
\begin{equation}
    \hat{H} = 
     \hat{H}_\mathrm{0} + \frac{\hbar\Omega}{2}\left(\hat{a}^\dagger\hat{\sigma}^- + \hat{a}\hat{\sigma}^+\right).
\label{eq:JC_lab_RWA}
\end{equation}
According to the Jaynes-Cummings Hamiltonian in Eq.~\eqref{eq:JC_lab_RWA}, where non-excitation-conserving transitions have been dropped by applying the RWA, we only have to consider excitation-conserving transitions between the qubit-resonator states $\ket{r, q}$: 
\begin{equation}
    \ket{n,g} \leftrightarrow \ket{n-1,e},
\end{equation}
where $n$ is the total number of excitations of the qubit-resonator state and the eigenvalue of the operator $\hat{N} = \ket{e}\bra{e} + \hat{a}^\dagger\hat{a}$. This operator commutes with the interaction term in the Hamiltonian:
\begin{equation}
    \begin{split}
        [\hat{H}_\mathrm{int},\hat{N}] = & \frac{\hbar\Omega}{2}\left( \hat{a}^\dagger[\hat{\sigma}^-,\ket{e}\bra{e}] + [\hat{a}^\dagger, \hat{a}^\dagger\hat{a}]\hat{\sigma}^- + \hat{a}[\hat{\sigma}^+,\ket{e}\bra{e}] + [\hat{a},\hat{a}^\dagger\hat{a}]\hat{\sigma}^+\right) \\
         = & \frac{\hbar\Omega}{2} \left( \hat{a}^\dagger \hat{\sigma}^- - \hat{a}^\dagger\hat{\sigma}^- - \hat{a} \hat{\sigma}^+ +\hat{a} \hat{\sigma}^+\right) = 0.
    \end{split}
\end{equation}
Using the eigenstates of $\hat{N}$ as a basis, the Hamiltonian can be represented by a block diagonal matrix:
\begin{equation}
    \hat{H}=\begin{bmatrix} 
1 &0 & 0 & 0&\cdots &\cdots &\cdots\\
0 & \hat{H}_1 & 0 & 0 &\ddots &\ddots &\ddots \\
0 & 0 & \hat{H}_2 & 0 & \ddots  & \ddots &\ddots \\
\vdots & \ddots   & \ddots &  \ddots &\ddots & \ddots & \ddots \\
\vdots &\ddots & \ddots & 0 & \hat{H}_n & 0 &\ddots \\
\vdots &\ddots&\ddots&\ddots&\ddots&\ddots & \ddots\\ 
\end{bmatrix}.
\end{equation}
The $(2\times2)$ matrices representing the manifolds with equal number of excitations $n$, are given by
\begin{equation}
\hat{H}_n = \begin{pmatrix}
\langle n-1,e|\hat{H}|n-1,e\rangle & \langle n-1,e|\hat{H}|n,g\rangle \\
 \langle n,g|\hat{H}|n-1,e\rangle &  \langle n,g|\hat{H}|n,g\rangle\\
\end{pmatrix} = \hbar
\begin{pmatrix}
\omega_\mathrm{r}(n-1)+ \frac{\omega_\mathrm{q}}{2} & \frac{\Omega\sqrt{n}}{2} \\
\frac{\Omega\sqrt{n}}{2} & n\omega_\mathrm{r} -\frac{\omega_\mathrm{q}}{2}\\
\end{pmatrix}.
\end{equation}
When reintroducing the zero-point energy of the computational resonator ($+\hat{I}\hbar\omega_\mathrm{r}/2$) in $\hat{H}_n$ and defining $\Delta = \omega_\mathrm{r}-\omega_\mathrm{q}$, we can rewrite the elements in the block-diagonal matrix as
\begin{equation}
    \hat{H}_n=\hbar\begin{pmatrix}
n\omega_\mathrm{r} - \frac{\Delta}{2} & \frac{\Omega\sqrt{n}}{2} \\
\frac{\Omega\sqrt{n}}{2} & n\omega_\mathrm{r} +\frac{\Delta}{2}\\
\end{pmatrix} = n\hbar\omega_\mathrm{r}\hat{I} - \frac{\hbar\Delta}{2}\pauliz + \frac{\hbar\Omega\sqrt{n}}{2}\paulix . \label{eq:Hn_lab}
\end{equation}
The time evolution in the lab frame can be calculated by the propagator
\begin{equation}
\begin{split}
    U_{n,\mathrm{lab}}(t,0) &=  e^{-i\hat{H}_nt/\hbar}\\
    &= e^{-in\omega_\mathrm{r}t}\begin{pmatrix}
        \mathrm{cos}\left(\frac{1}{2}\Omega(\Delta,n)t\right) + i\frac{ \Delta \cdot\mathrm{sin}\left(\frac{1}{2}\Omega(\Delta,n)t\right)}{\Omega(\Delta,n)} & -i\frac{ \Omega \sqrt{n}\cdot\mathrm{sin}\left(\frac{1}{2}\Omega(\Delta,n)t\right)}{\Omega(\Delta,n)}\\
        -i\frac{ \Omega \sqrt{n}\cdot\mathrm{sin}\left(\frac{1}{2}\Omega(\Delta,n)t\right)}{\Omega(\Delta,n)} & \mathrm{cos}\left(\frac{1}{2}\Omega(\Delta,n)t\right) - i\frac{ \Delta \cdot\mathrm{sin}\left(\frac{1}{2}\Omega(\Delta,n)t\right)}{\Omega(\Delta,n)} .
    \end{pmatrix}
\end{split}
\end{equation}
Here we have defined the photon number and detuning dependent generalised Rabi frequency $\Omega(\Delta,n) = \sqrt{\Delta^2 + n\Omega^2}$, which characterises the effective coupling between the states $|n-1,e\rangle$ and $|n,g\rangle$.
\paragraph{Free time evolution}
When the detuning between the qubit and the computational resonator is much larger than the Rabi frequency ($\sqrt{n}\Omega/\Delta \ll 1$), we can approximate $\Omega(\Delta,n) = \Delta\sqrt{1 + n\Omega^2/\Delta^2}\approx \Delta$. In this case, the unitary operator simplifies to the free time evolution:
\begin{equation}
    \begin{split}
        U_{n,\mathrm{lab}}(\sqrt{n}\Omega/\Delta \ll 1)
        \simeq 
         e^{-in\omega_\mathrm{r}t}\begin{pmatrix}
        e^{i\frac{\Delta}{2}t} & 0\\
        0 & e^{-i\frac{\Delta}{2}t} \label{eq:U_lab_Delta_inf_limit}
        \end{pmatrix}\\
    \end{split}
\end{equation}
\paragraph{Resonant population exchange}
To obtain a full population exchange between the states $|n-1,e\rangle$ and $|n,g\rangle$, the qubit and computational resonator have to be on resonance, i.e., $\Delta=0$ during the MOVE operation. Therefore the time evolution in the lab frame simplifies to
\begin{equation}
U_{n,\mathrm{lab}}(\Delta = 0) = e^{-in\omega_\mathrm{g}t}\begin{pmatrix}
        \mathrm{cos}\left(\frac{1}{2}\sqrt{n}\Omega t\right) & -i\mathrm{sin}\left(\frac{1}{2}\sqrt{n}\Omega t\right)\\
        -i\mathrm{sin}\left(\frac{1}{2}\sqrt{n}\Omega t\right) & \mathrm{cos}\left(\frac{1}{2}\sqrt{n}\Omega t\right)\end{pmatrix},\label{eq:time_evolution_lab_frame}
\end{equation}
where $\omega_\mathrm{g}$ is the gate frequency, meaning the frequency of both the qubit and the computational resonator during the Jaynes-Cummings gate. The Jaynes-Cummings gate is calibrated to achieve a full population exchange in the single excitation manifold, $n=1$. Assuming that $\Omega$ is constant during the gate, i.e., square flux pulse applied to the tunable coupler, the gate time follows as $t = \pi / \Omega$. By transforming from the lab frame to a rotating frame where both the qubit and the computational resonator are oscillating at $\omega_\mathrm{g}$ (gate frame), i.e., $\hat{R}_\mathrm{g} = e^{i n \omega_\mathrm{g}t \hat{I}}$, the time evolution for the lowest four excitation manifolds simplifies to 
\begin{equation}
    \begin{blockarray}{ccccccccc}
    \matindex{$\ket{0,g}$} & \matindex{$\ket{0,e}$} & \matindex{$\ket{1,g}$}  & \matindex{$\ket{1,e}$}  & \matindex{$\ket{2,g}$}  & \matindex{$\ket{2,e}$}  & \matindex{$\ket{3,g}$}  & \matindex{$\ket{3,e}$}  & \matindex{$\ket{4,g}$}\\
    \begin{block}{[ccccccccc]}
      1 & 0 & 0 & 0 & 0 & 0 & 0 & 0 & 0 \\
0 & 0 & -i & 0 & 0 & 0 & 0 & 0 & 0  \\
0 & -i & 0 & 0 & 0 & 0 & 0 & 0 & 0  \\
0 & 0 & 0 & \cos\left(\frac{\sqrt{2}}{2}\pi\right) & -i \sin\left(\frac{\sqrt{2}}{2}\pi\right) & 0 & 0 & 0& 0 \\
0 & 0 & 0 & -i \sin\left(\frac{\sqrt{2}}{2}\pi\right) & \cos\left(\frac{\sqrt{2}}{2}\pi\right) & 0 & 0 & 0 & 0  \\
0 & 0 & 0 & 0 & 0 & \cos\left(\frac{\sqrt{3}}{2}\pi\right) & -i  \sin\left(\frac{\sqrt{3}}{2}\pi\right) & 0 & 0  \\
0 & 0 & 0 & 0 & 0 & -i \sin\left(\frac{\sqrt{3}}{2}\pi\right) & \cos\left(\frac{\sqrt{3}}{2}\pi\right) & 0 & 0  \\
0 & 0 & 0 & 0 & 0 & 0 & 0 & -1 &  0 \\
0 & 0 & 0 & 0 & 0 & 0 & 0 & 0 & -1  \\
    \end{block}
  \end{blockarray}~.\label{eq:matrix_res_pop_exchange}
\end{equation}
In most cases, the rotating frame of choice is the so-called computational frame, where the free time evolution of all components is effectively fully compensated by the rotating frame, which means that the unitary transformation going from lab to the computational frame is constructed using $\hat{H}_0 = n\hbar\omega_\mathrm{r}\hat{I} - \frac{\hbar\Delta}{2}\pauliz$:
\begin{equation}
\hat{R}_\mathrm{lab\rightarrow comp} = e^{i\left(n\omega_\mathrm{r}\hat{I} -\frac{\Delta}{2}\pauliz\right)t} = e^{in\omega_\mathrm{r}t\hat{I}}e^{-i\frac{\Delta}{2}t\pauliz }.\label{eq:rotating_frame_comp}
\end{equation}
The Hamiltonian in the computational frame follows from Eq.~\eqref{eq:hamiltonian_rotatin_frame}
\begin{equation}
\begin{split}
    \hat{H}_\mathrm{n,comp} = \hat{R}_\mathrm{lab\rightarrow comp} \hat{V} \hat{R}_\mathrm{lab\rightarrow comp}^\dagger = \frac{\hbar \Omega \sqrt{n}}{2} e^{-i\frac{\Delta}{2}t\pauliz}  \paulix e^{i\frac{\Delta}{2}t\pauliz} = 
    \frac{\hbar \Omega \sqrt{n}}{2} \begin{pmatrix} 0 & e^{-i \Delta t}\\e^{i \Delta t} & 0\end{pmatrix}.
\end{split}
\end{equation}
The time evolution in the computational frame can be calculated using Eqs.~\eqref{eq:time_evolution_operator_transformation}, \eqref{eq:time_evolution_lab_frame} and \eqref{eq:rotating_frame_comp}:
\begin{equation}
\begin{split}
    U_{n,\mathrm{comp}}(t, t_0) &= \hat{R}_\mathrm{lab\rightarrow comp} U_{n,\mathrm{lab}}(t, t_0) \hat{R}_\mathrm{lab\rightarrow comp}^\dagger\\
    &= e^{in (\omega_\mathrm{r}-\omega_g)(t-t_0)}\begin{pmatrix}
        e^{-i\frac{\Delta}{2} (t-t_0)}\mathrm{cos}\left(\frac{\sqrt{n}\Omega}{2} (t-t_0)\right) & -ie^{-i\Delta \frac{t+t_0}{2}}\mathrm{sin}\left(\frac{\sqrt{n}\Omega}{2} (t-t_0)\right)\\
        -ie^{i\Delta \frac{t+t_0}{2}}\mathrm{sin}\left(\frac{\sqrt{n}\Omega}{2} (t-t_0)\right) & e^{i\frac{\Delta}{2} (t-t_0)}\mathrm{cos}\left(\frac{\sqrt{n}\Omega}{2} (t-t_0)\right)\end{pmatrix}.
\end{split}
\end{equation}
In the computational frame we see that the frame tracking need to take into account all the times when a Jaynes-Cummings gate is applied. Assuming the Jaynes-Cummings gate to be instantaneous ($t=t_0$) and calibrated to obtain a full population exchange in the one-excitation manifold, the time evolution operator simplifies to:
\begin{equation}
    U_{n,\mathrm{comp}}(t=t_0, t_0) = \begin{pmatrix}
        \mathrm{cos}\left(\frac{\sqrt{n}\pi}{2} \right) & -ie^{-i\Delta t}\mathrm{sin}\left(\frac{\sqrt{n}\pi}{2} \right)\\
        -ie^{i\Delta t}\mathrm{sin}\left(\frac{\sqrt{n}\pi}{2}\right) & \mathrm{cos}\left(\frac{\sqrt{n}\pi}{2}\right)\end{pmatrix}.
\end{equation}

\subsection{Populated Resonator Ramsey Experiment}
Next, we derive the time evolution for the populated Ramsey experiment, demonstrated in the main text, for arbitrary excitations. We change to obtained when changing to the reference frame defined by $\hat{H}_\mathrm{ref} = \hbar\omega_\mathrm{q} (\hat{a}^\dagger\hat{a} + \frac{\pauliz}{2})$. As we will see below, in this frame the time evolution adds phase factors proportional to the detuning $\Delta$ and the number of excitations in the resonator. Therefore when sweeping the wait time we will observe two frequency components in the excited state probability of the second move qubit, one at $f_1 = \Delta$ and another one at $f_2 = 2\Delta$, corresponding to one and two photons in the resonator, respectively. To transform the Hamiltonian into the this frame, we use the unitary
\begin{equation}
\hat{R}_\mathrm{1} = \quad e^{i\omega_\mathrm{q}t\left( {a}^\dagger{a}\hat{I} + \frac{\pauliz}{2}\right)}
= \quad e^{in\omega_\mathrm{q}t\hat{I}}e^{i\frac{\omega_\mathrm{q}}{2}t\pauliz }\\
= e^{in\omega_\mathrm{q}t} \begin{pmatrix}
e^{i{\omega_\mathrm{q}}t} &0\\
0&1
\end{pmatrix},\\
\label{eq:T2_frame_change}
\end{equation}
where we drop global phase terms. The second exponential adds an excitation-dependent phase to the corresponding state, while the matrix adds a qubit-state dependent phase. Using the considered reference frame Hamiltonian, the Jaynes-Cummings Hamiltonian can be rewritten as:
\begin{equation}
    \begin{split}
        \hat{H} = &\quad \hbar\omega_\mathrm{r}\hat{a}^\dagger\hat{a} + \hbar\omega_\mathrm{q}\frac{\pauliz}{2} + \frac{\hbar\Omega}{2}\left(\hat{a}^\dagger\hat{\sigma}^- + \hat{a}\hat{\sigma}^+\right)\\
        =&\quad \hbar\omega_\mathrm{q}\hat{a}^\dagger\hat{a} + \hbar\omega_\mathrm{q}\frac{\pauliz}{2} + \hbar(\omega_\mathrm{r}-\omega_\mathrm{q})\hat{a}^\dagger\hat{a} + \frac{\hbar\Omega}{2}\left(\hat{a}^\dagger\hat{\sigma}^- + \hat{a}\hat{\sigma}^+\right)\\
        =&\quad \hat{H}_\mathrm{ref} + \hbar\Delta\hat{a}^\dagger\hat{a} + \frac{\hbar\Omega}{2}\left(\hat{a}^\dagger\hat{\sigma}^- + \hat{a}\hat{\sigma}^+\right).
    \end{split}
\end{equation}
By applying Eq.~\eqref{eq:T2_frame_change}, the interaction Hamiltonian becomes:
\begin{equation}
        \hat{H}_\mathrm{int} = \quad \hbar n\Delta  \hat{I} -\frac{\hbar\Delta}{2}\pauliz+ \frac{\hbar\Omega\sqrt{n}}{2}\left[\mathrm{cos}(\omega_\mathrm{q}t) \paulix - \mathrm{sin}(\omega_\mathrm{q}t)\pauliy \right].
\end{equation}
We invoked the matrix representation of the Hamiltonian for a given excitation manifold $n$ and neglect the constant offset term which does not contribute to the dynamics of the system. The unitary operator in this frame becomes
\begin{equation}
    \begin{split}
        e^{-i\hat{H}_\mathrm{int}t/\hbar} = & \quad e^{-itn\Delta}e^{-i\frac{\Omega_\mathrm{eff}t}{2}\left(\frac{\Omega\sqrt{n}}{\Omega_\mathrm{eff}}\left(\mathrm{cos}(\omega_\mathrm{q}t) \paulix - \mathrm{sin}(\omega_\mathrm{q}t)\pauliy \right)-\frac{\Delta}{\Omega_\mathrm{eff}}\pauliz\right)}\\
        = & \quad e^{-itn\Delta} \renewcommand{\arraystretch}{2}\begin{pmatrix}
            \mathrm{cos}\left( \frac{\Omega_\mathrm{eff}t}{2} \right) + i\cdot \mathrm{sin}\left( \frac{\Omega_\mathrm{eff}t}{2} \right)\frac{\Delta}{\Omega_\mathrm{eff}} &  - i\cdot \mathrm{sin}\left( \frac{\Omega_\mathrm{eff}t}{2} \right)\frac{\Omega\sqrt{n}}{\Omega_\mathrm{eff}} e^{i\omega_\mathrm{q} t}\\
            - i\cdot \mathrm{sin}\left( \frac{\Omega_\mathrm{eff}t}{2} \right)\frac{\Omega\sqrt{n}}{\Omega_\mathrm{eff}} e^{-i\omega_\mathrm{q} t} & \mathrm{cos}\left( \frac{\Omega_\mathrm{eff}t}{2} \right) - i\cdot \mathrm{sin}\left( \frac{\Omega_\mathrm{eff}t}{2} \right)\frac{\Delta}{\Omega_\mathrm{eff}}
        \end{pmatrix},
    \end{split}
\end{equation}
where we define $\Omega_\mathrm{eff} = \sqrt{\Delta^2 +n \Omega^2}$. For free evolution when $\Omega\rightarrow 0$, we get
\begin{equation}
e^{-i\hat{H}_\mathrm{int}t/\hbar} = \quad e^{-itn\Delta} \renewcommand{\arraystretch}{2}\begin{pmatrix}
e^{it\frac{\Delta}{2}} &  0\\
0& e^{-it\frac{\Delta}{2}}
\end{pmatrix}
=\begin{pmatrix}
e^{-it(n-1)\Delta} &  0\\
0& e^{-itn\Delta}
\end{pmatrix}, \\
\end{equation}
where we neglect global phase factors. Considering the following bare eigenstates, their free time-evolution becomes
\begin{equation}
    \ket{0,e}  \rightarrow 1, \qquad
    \ket{1,e}  \rightarrow e^{-it\Delta}, \qquad
    \ket{2,g}  \rightarrow e^{-it2\Delta}.
\label{eq:H_ref_phases}
\end{equation}
The state evolution with the first qubit and computational resonator right after the MOVE derived from the circuit can now be written as
\begin{equation}
    \ket{0,g} \rightarrow -i\ket{0,e} \rightarrow -\ket{1,g}. 
\end{equation}
Since $-1$ is a global phase, we can neglect it. In the following, we use the previously introduced shorthand notation $c_2 = \mathrm{cos} \left(\frac{\sqrt{2}\pi}{2}\right)$ and $s_2 = \mathrm{sin} \left(\frac{\sqrt{2}\pi}{2}\right)$. In the subsequent part of the circuit up to the second MOVE, the second qubit and computational resonator state becomes
\begin{equation}
    \ket{1,g}\rightarrow\frac{\ket{1,g}-i\ket{1,e}}{\sqrt{2}} \rightarrow  \frac{1}{\sqrt{2}} \left( -i\ket{0,e} -i \left( c_2 \ket{1,e} -i s_2 \ket{2,g}  \right)\right) 
    =  \frac{-i}{\sqrt{2}} \left(\ket{0,e} + c_2 \ket{1,e} - i s_2 \ket{2,g}\right).
\end{equation}
In the reference frame, when the qubit and resonator are far detuned and the interaction term is considered to be negligible, we see from Eqs.~\eqref{eq:H_ref_phases} that the free time-evolution of the state becomes
\begin{equation}
        e^{-i\hat{H}t/\hbar}\ket{\psi} = \frac{1}{\sqrt{2}}\left(\ket{0,e} + c_2 \ket{1,e} e^{-it\Delta}- i\cdot s_2 \ket{2,g}e^{-it2\Delta}\right).
\end{equation}
After the second MOVE gate, we have
\begin{align}
    \frac{-i}{\sqrt{2}} \bigg(&  \ket{1,g}
    + i \ket{1,e} \left( (c_2)^2 e^{-it\Delta} - (s_2)^2 e^{-it2\Delta}\right)
    + \ket{2,g} c_2s_2 \left( e^{-it\Delta} + e^{-it2\Delta} \right) \bigg).
\end{align}
After the final $\frac{\pi}{2}$-pulse we obtain the state
\begin{align}
 \frac{1}{2} \bigg(\ket{1,g} \left(1 + (c_2)^2 e^{-it\Delta} -  (s_2)^2 e^{-it2\Delta}\right)
    & - i \ket{1,e} \left(1 - (c_2)^2 e^{-it\Delta} + (s_2)^2 e^{-it2\Delta}\right) \\ &
    + \ket{2,g} c_2s_2 \left( e^{-it\Delta} + e^{-it2\Delta} \right)
    -i\ket{2,e} c_2s_2 \left( e^{-it\Delta} + e^{-it2\Delta} \right) \bigg).
\end{align}
The final probabilities of the qubit being in the excited state can be written as
\begin{equation}
    P_e = \sum_{n=0}^\infty |\braket{n, e|\psi}|^2 =  \braket{\psi|1, e}\braket{1, e|\psi} + \braket{\psi|2, e}\braket{2, e|\psi}.\label{eq:Pe_general}
\end{equation}
The last equality follows from the fact that we only consider the first and the second excitation manifold. These terms are
\begin{equation}
        P_{1e} =  |\braket{\psi|1, e}|^2 
        = \frac{1}{4}\left[1 - 2(c_2)^2 \cos(t\Delta) + 2(s_2)^2 \cos(t2\Delta) - 2(c_2)^2 (s_2)^2 \cos(t\Delta) + (c_2)^4 + (c_2)^4\right],
\end{equation}
\begin{align}
        P_{2e} = |\braket{\psi|2, e}|^2
        = \frac{(c_2s_2)^2}{2}(1 +\mathrm{cos}(t\Delta)).\\
\end{align}
Taking the sum of these equations, the full qubit excitation becomes
\begin{equation}
    P_{e} = \frac{1}{2}\left( 1 -c_2^2 \cos(\Delta t) + s_2^2 \cos(2\Delta t)\right).
\end{equation}
As an outlook, we generalize the populated Ramsey experiment to arbitrary resonator states. For simplicity, we assume ideal JC interaction without any additional phase shifts here, the calculation for the imperfect JC interaction is analogous but requires a bit more book keeping. We assume the initial state 
\begin{equation}
|\psi\rangle = \sum_n \alpha_n |g,n\rangle.
\end{equation}
Applying the first $\pi/2$-pulse, followed by the first MOVE gate in the Ramsey sequence leads to
\begin{equation}
|\psi\rangle = \frac{1}{\sqrt{2}}\sum_n \alpha_n \bigg \{ c_n |g,n\rangle - i s_n |e,n-1 \rangle -i c_{n+1}|e,n\rangle - s_{n+1}|g,n+1\rangle \bigg \}.
\end{equation}
Subsequent time evolution in the lab frame is
\begin{equation}
|g,n \rangle \to |g,n \rangle e^{i n \omega_\mathrm{r} t}, \qquad |e,n \rangle \to |g,n \rangle e^{i n \omega_\mathrm{r} t + i\omega_\mathrm{q} t}.
\end{equation}
After the final MOVE gate, the state can be written as
\begin{equation}
|\psi\rangle = \frac{1}{\sqrt{2}}\sum_n \alpha_n \bigg \{\beta_n |g,n\rangle + \gamma_n|e,n\rangle + \lambda_n|e,n-1\rangle + \kappa_n|g,n+1 \rangle  \bigg \},
\end{equation}
where
\begin{equation}
\beta_n \equiv c_n^2 e^{i n \omega_\mathrm{r} t} - s_n^2 e^{i(n - 1)\omega_\mathrm{r}t + i \omega_\mathrm{q}t}, \qquad \gamma_n \equiv -i \left[c_{n+1}^2 e^{i n \omega_\mathrm{r} t + i \omega_\mathrm{q}} - s_{n+1}^2 e^{i(n+1)\omega_\mathrm{r}t}\right]
\end{equation}
\begin{equation}
\lambda_n \equiv -i c_n s_n \left[e^{i(n-1)\omega_\mathrm{r} t + i\omega_\mathrm{q}t} + e^{i n \omega_\mathrm{r} t}\right], \qquad \kappa_n \equiv -c_{n+1}s_{n+1}\left[e^{i n \omega_\mathrm{r} t + i\omega_\mathrm{q} t} + e^{i (n+1)\omega_\mathrm{r}t} \right].
\end{equation}
The final state contributions with the qubit in the excited state are
\begin{equation}
\langle e|\psi\rangle = \frac{1}{2}\sum_n \alpha_n \bigg \{(\gamma_n - i \beta_n)|n\rangle + \lambda_n |n-1\rangle + \kappa_n |n+1\rangle \bigg \}.
\end{equation}
Thus, the excited state probability is
\begin{equation}
P_\mathrm{e} = \frac{1}{4}\sum_n |\alpha_n|^2 \bigg \{ |\gamma_n - i \beta_n|^2 + |\lambda_n|^2 + |\kappa_n|^2 \bigg \}.
\end{equation}
The individual probabilities can be expressed as
\begin{align}
|\gamma_n - i \beta_n|^2 &= c_n^4 + s_n^4 + c_{n+1}^4 + s_{n+1}^4 - 2c_{n+1}^2 s_{n+1}^2 \cos((\omega_\mathrm{r}-\omega_\mathrm{r})t) \\ 
& + 2c_n^2 c_{n+1}^2 \cos(\omega_\mathrm{q}t) - 2c_{n+1}^2 s_n^2 \cos(\omega_\mathrm{r}t) - 2c_n^2 s_{n+1}^2 \cos(\omega_\mathrm{r}t ) \\
& -2s_s^2 s_{n+1}^2 \cos((2\omega_\mathrm{r} - \omega_\mathrm{q})t) - 2c_n^2 s_n^2 \cos((\omega_\mathrm{r} - \omega_\mathrm{q})t),
\end{align}
\begin{equation}
|\lambda_n|^2 \equiv 2c_n^2 s_n^2 \left[1 + \cos((\omega_\mathrm{r} - \omega_\mathrm{q})t) \right], \qquad
|\kappa_n|^2 \equiv 2c_{n+1}^2 s_{n+1}^2 \left[1 + \cos((\omega_\mathrm{r} - \omega_\mathrm{q})t) \right].
\end{equation}
Note that just as for the two-photon case, there are only second harmonic contributions $2\omega_\mathrm{r}$ from the resonator. The reason is the equidistant spacing of the harmonic oscillator combined with the $2\times2$ block diagonal matrix structure of the JC Hamiltonian, implying that all rotating contributions spaced apart further than $2\omega_\mathrm{r}$ act as a global phase for the respective state probabilities and thus drop out. The final result in the lab frame is given by
\begin{equation}
P_e = \frac{1}{2}\sum_n |\alpha_n|^2 \bigg \{1 + c_n^2 s_n^2 \cos(\omega_\mathrm{q}t) -(c_{n+1}^2 s_n^2 + c_n^2 s_{n+1}^2)\cos(\omega_\mathrm{r}t) + s_n^2 s_{n+1}^2\cos((2\omega_\mathrm{r} - \omega_\mathrm{q})t)\bigg \}.
\end{equation}
And the detected probability (assuming virtual detuning $\Omega = 0$) is 
\begin{equation}
P_e = \frac{1}{2}\sum_n |\alpha_n|^2 \bigg \{1 + c_n^2 s_n^2 -(c_{n+1}^2 s_n^2 + c_n^2 s_{n+1}^2)\cos(\Delta t) + s_n^2 s_{n+1}^2\cos(2\Delta t)\bigg \}.
\end{equation}
In the case $n=0$, this expression reproduces the standard Ramsey oscillations and for $n=1$, we reproduce Eq.~\eqref{Eq:JC_probability_with_phase_shift} without any phase shifts.

\subsection{Calibration of the Jaynes-Cummings gate phase corrections}
In this section, we explain how the populated Ramsey experiment can be used to calibrate the single-qubit phase corrections appearing in the unitary of the JC gate for up to two excitations in the qubit-resonator system.
In a realistic experimental setting, the JC interaction is of the form of a general excitation-preserving exchange. To use the JC gate for actual algorithm applications, the phases $(\gamma_n, \chi_n, \zeta_n)$ appearing in Eq.~\eqref{Eq:JC_probability_with_phase_shift}, need to be calibrated for each $n$-photon manifold of the block-diagonal JC time-evolution operator. Since we apply the MOVE operation only in pairs and assume the MOVE to perform a full state transfer in the $n=1$ manifold, the phases $\chi_1$ and $\zeta_1$ are irrelevant here. In the single photon manifold, only the common phase $\gamma_1$ needs to be calibrated and is corrected for using a VZ gate as explained in Appendix~\ref{Sec:MOVE_fine_VZ}. Therefore, we set $\gamma_1 = 0$ in the following, as we compensate this non-ideal phase in the experiment sequence. The populated Ramsey experiment can then be used to calibrate the phases for the two-photon manifold. The state after the second MOVE gate in the populated resonator Ramsey sequence can be expressed as
\begin{equation}
\frac{i}{\sqrt{2}}\left(|1,g\rangle e^{i\omega_\mathrm{q}t} +a|1,e\rangle + b|2,g\rangle \right),
\end{equation}
where we have
\begin{equation}
a \equiv i\left(c_+^2 e^{i(\omega_\mathrm{q} + \omega_\mathrm{r})t} - s_- s_+ e^{2i \omega_\mathrm{r}t}\right), \qquad b \equiv c_+ s_- e^{i(\omega_\mathrm{q} + \omega_\mathrm{r})t} + c_-s_- e^{2i\omega_\mathrm{r}t},
\end{equation}
with the abbreviation $c_\pm, s_\pm \equiv c_2^{\pm}, s_2^{\pm}$ [cf.\,Eq.~\eqref{Eq:definition_cnpm_snpm}]. In this case, we find 
\begin{equation}
P_{1e} = \frac{1}{4}\left[1 + c_2^4 + s_2^4 - 2c_2^2 \cos(\omega_\mathrm{r}t -2\gamma_2 - 2\zeta_2) + 2s_2^2 \cos(2\omega_\mathrm{r}t - \omega_\mathrm{q}t -2\gamma_2) - 2c_2^2s_2^2 \cos(\omega_\mathrm{r}t - \omega_\mathrm{q}t + 2\zeta_2)\right]
\end{equation}
\begin{equation}
P_{2e} = \frac{c_2^2 s_2^2}{2}\left[1 + \cos((\omega_\mathrm{r} - \omega_\mathrm{q})t + 2\zeta_2)\right].
\end{equation}
Also in this case, the Ramsey sequence filters the terms rotating with frequency $\omega_\mathrm{r} - \omega_\mathrm{q}$. The detected excited state probability in the qubit rotating frame is 
\begin{equation}\label{Eq:probability_with_phase_shift}
P_e = \frac{1}{2}\left[1 - c_2^2 \cos(\Delta t -2\gamma_2 - 2\zeta_2) + s_2^2 \cos(2\Delta t  -2\gamma_2) \right],
\end{equation}
i.e., the non-calibrated phases in the JC interaction directly manifest as phase shifts of the detected populated Ramsey interference pattern. To restore the ideal interference pattern, one can compensate for these phase shifts by introducing additional control degrees of freedom. 
Since the relative phase $\chi_2$ cancels out when applying an even number of MOVE operations, only two degrees of freedom, $\gamma_2$ and $\zeta_2$, are required in Eq.~\eqref{Eq:probability_with_phase_shift}. In the simplest case, those can be included in the form of subsequent physical $z$-rotations with phases $\varphi$ and $\varphi^\prime$ after the two MOVE operations in the Ramsey sequence, as shown in Fig.~\ref{Fig:Fig_JC}(b).

For this modified circuit, the state after the first MOVE gate in the Ramsey sequence transforms to
\begin{equation}
-\frac{1}{\sqrt{2}}\left(|0,e\rangle e^{i\varphi} + c_+ |1,e\rangle e^{i\varphi} - i s_- |2,g\rangle\right),
\end{equation}
and after the second MOVE gate
\begin{equation}
\frac{i}{\sqrt{2}}\left(|1,g\rangle e^{i\omega_\mathrm{q}t + i\varphi} +a|1,e\rangle + b|2,g\rangle \right),
\end{equation}
where we have to redefine $a$ and $b$ as
\begin{equation}
a \equiv i\left(c_+^2 e^{i(\omega_\mathrm{q} + \omega_\mathrm{r})t + i\varphi^\prime} - s_- s_+ e^{2i \omega_\mathrm{r}t + i(\varphi - \varphi^\prime)}\right), \qquad b \equiv c_+ s_- e^{i(\omega_\mathrm{q} + \omega_\mathrm{r})t +i\varphi} + c_-s_- e^{2i\omega_\mathrm{r}t}.
\end{equation}
The final excited state probability is given by
\begin{equation}
P_e = \frac{1}{2}\left[1 - c_2^2 \cos(\Delta t -2\gamma_2 - 2\zeta_2 +\varphi^\prime) + s_2^2 \cos(2\Delta t -2\gamma_2 + \varphi^\prime - \varphi) \right],
\end{equation}
and therefore, we can restore the ideal Jaynes-Cummings evolution by applying the following phase corrections
\begin{equation}
 \varphi = 2\zeta_2, \qquad \varphi^\prime = 2\gamma_2 + 2\zeta_2.
\end{equation}

\medskip\noindent

\clearpage

\end{document}